\documentclass[a4paper,11pt]{article}
\pdfoutput=1 

\usepackage{jheppub} 

\usepackage[T1]{fontenc} 
\usepackage[multiple]{footmisc}
\usepackage{booktabs}
\usepackage{multirow}
\usepackage{pbox}
\usepackage{tabularx}
\usepackage{slashed}
\usepackage{snapshot}

\renewcommand{\arraystretch}{1.5}
\newcommand{\newc}{\newcommand}
\def\beq{\begin{equation}}
\def\eeq{\end{equation}}
\def\beqn{\begin{eqnarray}}
\def\eeqn{\end{eqnarray}}
\newc{\gluino}{\tilde g}
\newc{\Conep}{\tilde \chi_1^+}
\newc{\Conem}{\tilde \chi_1^-}
\newc{\Cone}{\tilde \chi_1^\pm}
\newc{\Ntwo}{\tilde \chi_2^0}
\newc{\None}{\tilde \chi_1^0}
\newcommand{\lsim}{\raisebox{-0.13cm}{~\shortstack{$<$ \\[-0.07cm] $\sim$}}~} 
\newcommand{\gsim}{\raisebox{-0.13cm}{~\shortstack{$>$ \\[-0.07cm] $\sim$}}~} 

\def\HWPP{\textsf{HERWIG++}~}

\preprint{\parbox{4cm}{ZU-TH 10/14\\MCnet-14-07\\LPN14-053\\KCL-PH-TH/2014-08\\LCTS/2014-10}}

\title{
Higgs boson to di-tau channel in Chargino-Neutralino searches at the LHC
}

\author[a]{Andreas Papaefstathiou,}
\author[b]{Kazuki Sakurai,}
\author[c]{Michihisa Takeuchi}

\affiliation[a]{Physik Institut, Universit\"at
  Z\"urich, Switzerland}
\affiliation[b,c]{Department of Physics, Theoretical Particle Physics \&
  Cosmology, King's College London, United Kingdom.}
\emailAdd{andreasp@physik.uzh.ch}
\emailAdd{kazuki.sakurai@kcl.ac.uk}
\emailAdd{michihisa.takeuchi@kcl.ac.uk}

\abstract{We consider chargino-neutralino production, $\Ntwo
\Cone \to (h \None )(W^\pm \None)$, which results in Higgs boson final
  states that subsequently decay (inclusively) to leptons (either
  $h\to \tau^+\tau^-$ or $h\to W^+W^- \to (e^+e^-, \mu^+
  \mu^-, \tau^+\tau^-)+\slashed{E}_T$). Such channels are dominant
  in large regions of the allowed supersymmetric parameter space for many concrete
  supersymmetric models. The existence of leptons allows
  for good control over the backgrounds, rendering this channel
  competitive to the conventional $h\to
  b\bar{b}$ channel that has been previously used to impose constraints. We include
  hadronic decays of the $\tau$ leptons in our analysis through a
  $\tau$-identification algorithm. We consider integrated
  luminosities of 100~fb$^{-1}$, 300~fb$^{-1}$ and 3000~fb$^{-1}$, for an LHC running at
  $pp$ centre-of-mass energy of 14~TeV and provide the expected constraints on the
  $M_2$-$M_1$ plane.}

\begin{document} 
\maketitle
\flushbottom

\section{Introduction}\label{sec:intro}

One of the primary goals of the CERN Large Hadron Collider (LHC) is to discover or rule out weak scale supersymmetry (SUSY).
So far the ATLAS and CMS collaborations have conducted a number of direct SUSY searches in many different channels.
The absence of excesses in those channels over the Standard Model (SM) background in turn placed impressive constraints on the SUSY parameter space. 
The limit is particularly stringent for coloured SUSY particles because of their large production cross sections.  
For instance, gluino and light flavour squarks are excluded up to
masses of about $1 - 1.5$~TeV~\cite{TheATLAScollaboration:2013fha,TheATLAScollaboration:2013uha,
TheATLAScollaboration:2013via,
Aad:2013wta,
CMS:2013cfa,
CMS:2014ksa,
Chatrchyan:2014lfa}, 
although the precise mass bounds depend on the details of the decay chains and mass spectrum.~\footnote{
See, {\it e.g.}, \cite{fastlim, Kraml:2013mwa, Conte:2012fm, Drees:2013wra} for the recent ideas and programmes to address this problem. 
} 

The recent observation of a SM-like Higgs boson~\cite{ATLAS_Higgs, CMS_Higgs}
also provides interesting implications and opportunities for the exploration of SUSY phenomenology.
First of all, the observed mass $\sim 125$\,GeV and the measured properties of the SM-like Higgs boson are 
consistent with the lightest CP-even Higgs ($h$) in the minimal SUSY extension of the SM (MSSM)
especially when the masses of scalar superparticles are larger than the several TeV~\cite{Giudice:2011cg, Kane:2011kj, Ibe:2011aa, Djouadi:2013vqa}.
Such scenarios are also consistent with the null results of direct
SUSY searches and the precise measurements of flavour-changing neutral
currents (FCNC) and CP-violating observables. 

Even though the scalars are anticipated to be heavy, it is possible to have relatively light gauginos in the SUSY spectrum.
In particular, the electorweak (EW) gauginos can exist and still be
very light, since their production cross sections are much smaller than
the coloured SUSY particles of the same mass. Indeed, in concrete models, the EW gauginos tend to be much lighter than the coloured SUSY particles.
This is due to the fact that the renormalisation group evolution (RGE)
increases coloured SUSY particle masses at low energies, owing to
their strong QCD interaction, whilst the effect is much smaller for EW gauginos.
It is known~\cite{Baer:2012ts} that if the gaugino GUT relation ($M_3:M_2:M_1 \sim 7:2:1$) holds, 
the production of EW gauginos can dominate over gluino pair production at
the 14 TeV LHC due to the mass hierarchy. Moreover, many SUSY breaking scenarios predict a large mass splitting between gauginos and scalars~\cite{Giudice:1998xp, ArkaniHamed:2004fb, Acharya:2008bk, Arvanitaki:2012ps, Hall:2011jd, Ibe:2012hu}.
Unlike the scalar masses, gaugino mass terms are prohibited by R-symmetry,
and their mass generation mechanism may be very different.
In the scenarios where R-symmetry is only weakly broken, the gauginos tend to be much lighter than the scalars.
In such scenarios, gauginos are the only SUSY particles which are accessible at the LHC~\cite{
Wells:2003tf, Bhattacherjee:2012ed, Hall:2012zp, ArkaniHamed:2012gw}.

The EW gauginos, namely, charginos and neutralinos, have already been intensively searched for at the LHC.
ATLAS and CMS interpreted their results in the context of simplified
models, where several assumptions were made. For instance, the
lightest neutralino ($\None$) was assumed to be bino-like and the
second lightest neutralino ($\Ntwo$) and the lighter chargino
($\Cone$) wino-like, while possessing the same mass, $m_{\Ntwo} = m_{\Cone}$.
In these simplified models, particular decays of the chargino and the
second lightest neutralino with 100\,\% branching ratios were considered.
The most stringent constraints were found for the models where the $\Ntwo$ and $\Cone$ decay exclusively into on-shell sleptons ($\tilde \ell$ and $\tilde \nu$).  
In this case, $m_{\Ntwo} = m_{\Cone}$ is excluded up to about 700~GeV
with $m_{\None} \lsim 300$~GeV~\cite{Aad:2014nua, CMS:2013dea}. Simplified models with the chargino and neutralino decays leading to di-$\tau$ final states via on-shell $\tilde \tau$ and $\tilde \nu_\tau$ have also been searched for,
and the limit was found to be $m_{\Ntwo} = m_{\Cone} \gsim 300$ (350) GeV with $m_{\None} \lsim 100$ (50) GeV~\cite{CMS:2013dea, Aad:2014nua}. 
If the sleptons and staus are heavier than the EW gauginos, the $\Cone$ predominantly decays to $W^\pm$ and $\None$.
On the other hand, the $\Ntwo$ has two possible decay modes: $\Ntwo \to Z \None$ and $\Ntwo \to h \None$.   
The former has been searched for and the resulting limit was $m_{\Ntwo} = m_{\Cone} \gsim 350$ 
with $m_{\None} \lsim 100$ GeV~\cite{Aad:2014nua}.
The latter process, $\Ntwo \Cone \to (h \None)(W^\pm \None)$, has also
been looked for recently by employing the $h \to b \bar b$ channel.
This channel suffers from an overwhelmingly large $t \bar t$
background  and only weak constraints have been found.
The bound is $m_{\Ntwo} = m_{\Cone} \gsim 200$~GeV~\cite{CMS:2013afa} and 300~GeV~\cite{TheATLAScollaboration:2013zia} only when $m_{\None} \lsim 30$ GeV.

The fact that current searches provide weak constraints is not the
only reason the $\Ntwo \Cone \to (h \None)(W^\pm \None)$ process is
especially interesting for further study. Firstly, in this process one can take advantage of the discovery of
the SM-like Higgs boson, making use of the of its properties as measured in the present dataset~\cite{Han:2013kza}.
Identifying the observed boson as the lightest CP-even Higgs in the MSSM
allows us to make a precise prediction of the $\Ntwo \Cone \to (h \None)(W^\pm \None)$ signature,
which is necessary for the limit calculation and also useful in
designing optimal search strategies for this mode.
Secondly, as we will see in Section~\ref{sec:mode}, 
the scenarios with heavy scalar SUSY particles
may imply that $\Ntwo$ predominantly decays into $h$ and $\None$.

In this paper, we study the exclusion and discovery reach of the $\Ntwo \Cone \to (h \None)(W^\pm \None)$ process,
using the $W$ decays to electrons muons or taus and the $h \to \tau \tau$ and $h \to W W \to (\tau/\ell, \nu)(\tau/\ell, \nu)$ modes.
Our study differs from earlier studies for $\Ntwo \Cone \to (h \None)(W^\pm \None)$~\cite{Baer:2012ts, Han:2013kza, Ghosh:2012mc}, 
which have focused on the decays of the $W$ to electrons or muons and $h \to b \bar b$ modes.
The obvious advantage of the channel with $h \to b \bar b$ is its
relatively large branching ratio BR($h \to b \bar b$).
However, this channel suffers from an overwhelming $t \bar t$ background.
Employing $h \to \tau \tau$ and $h \to W W \to (\tau/\ell, \nu)(\tau/\ell, \nu)$ introduces a reduction of the branching ratio,
by a factor of $[BR(h \to \tau \bar \tau) + BR(h \to W W \to (\tau/\ell, \nu)(\tau/\ell, \nu))] / BR(h \to b \bar b) \sim 0.15$,
but the $t \bar t$ background can be reduced significantly by vetoing $b$-jets and requiring two $\tau$s in the final state as we will see in Section~\ref{sec:analysis}. 
We will demonstrate that the channel with $h \to \tau\tau$ and $h \to W W \to
(\tau/\ell, \nu)(\tau/\ell, \nu)$ can provide competitive discovery
and exclusion prospects to those obtained in the channel with the $h \to b \bar b$ mode.

Revealing the details of the EW gaugino sector is especially
important. It is commonly believed that this sector contains the particle
that can be a candidate for dark matter. Moreover, studying the accessible mass scale of the EW gauginos at the LHC is important \cite{Bharucha:2014ama} for the planning of future collider programmes.
  
The article is organised as follows: in the next section, we provide
the details of the setup we use for the $\Ntwo \Cone \to (h \None)
(W^\pm \None)$ mode and discuss the cross section and branching ratio of EW gauginos with
particular attention to heavy scalar scenarios with a large $\mu$-term. In Section~\ref{sec:analysis} we
provide details of the Monte Carlo simulation performed to generate
the samples used in the analysis and give details of the algorithm
employed for the identification of jets originating from hadronic decays of
$\tau$ leptons. We then proceed to outline the details of our discrimination analysis, which forms the basis for defining the signal regions for
the SUSY parameter space scan. The results of the parameter space scan
are presented and discussed in Section~\ref{sec:results}. We conclude
in Section~\ref{sec:conclusions}. 
Supplementary appendices describe
the definition of a kinematic variable used in our analysis, the
calculation of cross sections for signal and background and
statistical methods with low event numbers. The last appendix in particular describes a systematic way to recast our results onto the other scenarios.
The application includes higgsino NSLP scenarios with a bino LSP and
higgsino/wino NLSP scenarios with a gravitino LSP as discussed, for example, in \cite{Asano:2010ut, Kats:2011qh, Curtin:2012nn, Hikasa:2014yra}.   

\section{The $\Ntwo \Cone \to (h \None) (W^\pm \None)$ mode}\label{sec:mode}

In this section we describe the setup of our analysis and clarify the assumptions we made in the chargino and neutralino sectors.  
Moreover, we discuss the cross sections and branching ratios of the
production and decay modes relevant to our analysis.

\subsection{The setup}\label{sec:setup}

Throughout this paper we consider CP-conserving EW gaugino sector and assume
$ m_{\tilde{\chi}_2^0} \simeq m_{\tilde{\chi}_1^\pm} > m_{\tilde{\chi}_1^0}$ for simplicity.
This relation is realised in many SUSY breaking scenarios, 
particularly 
in the cases where
$|\mu| \gg M_2 > M_1$ and $M_2 \gg |\mu| > M_1$.
The former case is motivated by the heavy scalar scenario.
In the MSSM, the soft scalar masses for $H_u$ and $H_d$ and the $\mu$-parameter are related by the EW symmetry breaking condition~\cite{primer}
\beqn
\frac{m_Z^2}{2} =  \frac{m_{H_d}^2 - m_{H_u}^2 \tan^2 \beta}{ \tan^2 \beta - 1}  - |\mu|^2  \,.
\label{eq:mu}
\eeqn
This condition implies that the $\mu$-parameter is expected to be of
the same scale as the scalar masses, unless $m_{H_u}$ and $m_{H_d}$ are carefully tuned at the EW scale in such a way that the first terms in 
the right hand side of Eq.~(\ref{eq:mu}) becomes unnaturally small.\footnote{
Even in that case, the same size of tuning is required on the $\mu$-parameter. 
}

In this section we assume the scale of $\mu$ is equal to the scalar masses and $|\mu| \gg M_2 > M_1 > 0$.
However, the collider analysis described in Section~\ref{sec:analysis}
is applicable to other scenarios as far as the $\tilde N \tilde C^{\pm} \to (h \chi) (W^\pm \chi)$ topology is concerned,
where $\tilde N$ and $\tilde C^{\pm}$ are massive BSM particles with the same mass and $\chi$ is an invisible particle with an arbitrary mass.
One such scenario involves a bino LSP scenario with a higgsino NLSP, $M_2 \gg |\mu| > M_1$.
The application also includes gravitino LSP scenarios with wino or higgsino NSLP 
as discussed for example in~\cite{Asano:2010ut, Kats:2011qh, Curtin:2012nn, Hikasa:2014yra},
where the same topology is realised by $\tilde \chi_1^0 \tilde \chi_1^\pm \to (h \tilde G) (W^\pm \tilde G)$ with $\tilde G$ being gravitino.
We will get back to this point in the end of this section.

\subsection{The cross sections}\label{sec:cross}

\begin{figure}[!t]
  \begin{center}
    \includegraphics[clip, scale=1.]{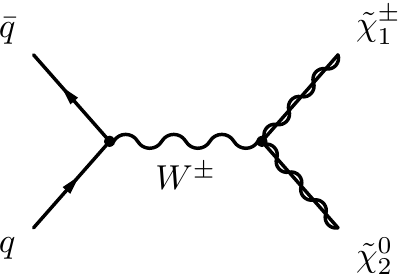}
    \hspace{3mm}
    \includegraphics[clip, scale=1.]{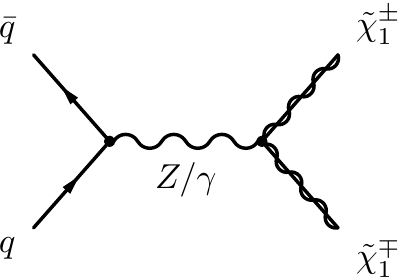}    
    \hspace{3mm}
    \includegraphics[clip, scale=1.]{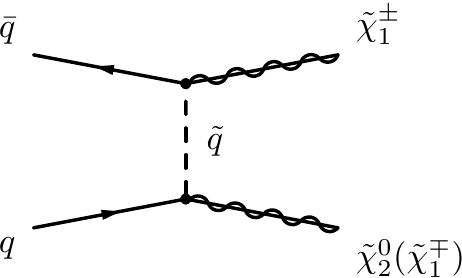}        
  \end{center}
  \caption{The tree-level diagrams for the relevant $\Ntwo$ and $\Cone$ production.}
  \label{fig:diagram}
\end{figure}

Fig.~\ref{fig:diagram} shows the tree-level diagrams for the relevant
modes of $\Ntwo$ and $\Cone$ production.
There are two types of diagrams which may interfere: $s$-channel diagrams with gauge boson exchange and $t$-channel diagrams with
squark exchange.  
The $t$-channel diagrams are suppressed by the squark mass and it is
expected that the contribution of this diagram decreases as the squark
mass increases.

\begin{figure}[!t]
	\begin{center}
          \includegraphics[scale=0.55]{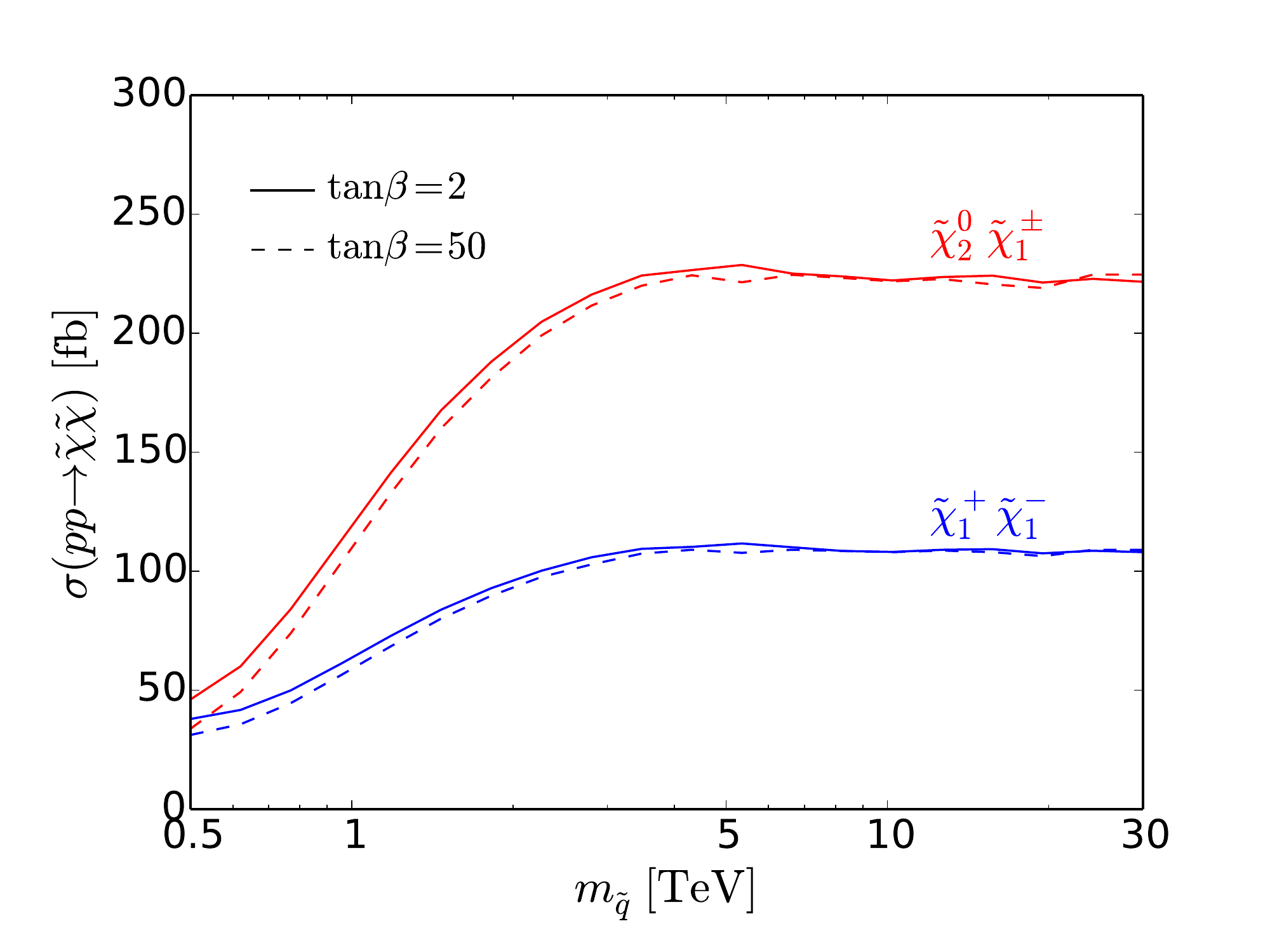}
	\caption{The NLO production cross sections for the $\None
          \Cone$ and $\tilde \chi_1^+ \tilde \chi_1^-$ modes at the 14 TeV LHC as
          functions of the squark mass. The cross sections have been
          calculated using {\tt Prospino 2.1}~\cite{Beenakker:1996ed,
            Beenakker:1999xh} with all the charges summed. We have set $\mu = m_{\tilde q}$ and
          $M_2 =350$~GeV and $M_1 = 100$~GeV. The solid and dashed
          curves correspond to $\tan \beta = 2$ and 50,
          respectively. }
	\label{fig:xsec}
	\end{center}
\end{figure}

Fig.~\ref{fig:xsec} shows the NLO production cross sections for the
$\Ntwo \Cone$ and $\tilde \chi_1^+ \tilde \chi_1^-$ modes at the
14~TeV LHC as functions of the squark mass.
The cross sections have been calculated using {\tt Prospino
  2.1}~\cite{Beenakker:1996ed, Beenakker:1999xh} with all the charges summed.
In the plot and throughout the paper, we take $|\mu| = m_{\tilde q}$ for simplicity. 
For the specific plot, we take $M_2 = 350$~GeV and $M_1 = 100$~GeV.
The solid and dashed curves correspond to $\tan \beta = 2$ and 50,
respectively. As a result of destructive interference between the $s$-channel gauge boson exchange diagram and the $t$--channel squark 
exchange diagram, the $\Ntwo \Cone$ and $\tilde \chi_1^+ \tilde
\chi_1^-$ production cross sections increase as the squark mass increases.
For a squark mass larger than $\sim 4$~TeV, the contribution of the squark exchange diagram is decoupled
and the cross sections become insensitive to the squark mass.
It is interesting to note that the $\Ntwo \Cone$ and $\tilde \chi_1^+ \tilde \chi_1^-$ cross sections are maximised in the limit of large squark mass.
This gives additional motivation to perform EW gaugino searches in the
context of heavy scalar scenarios.

\begin{figure}[!t]
	\begin{center}
          \includegraphics[scale=0.55]{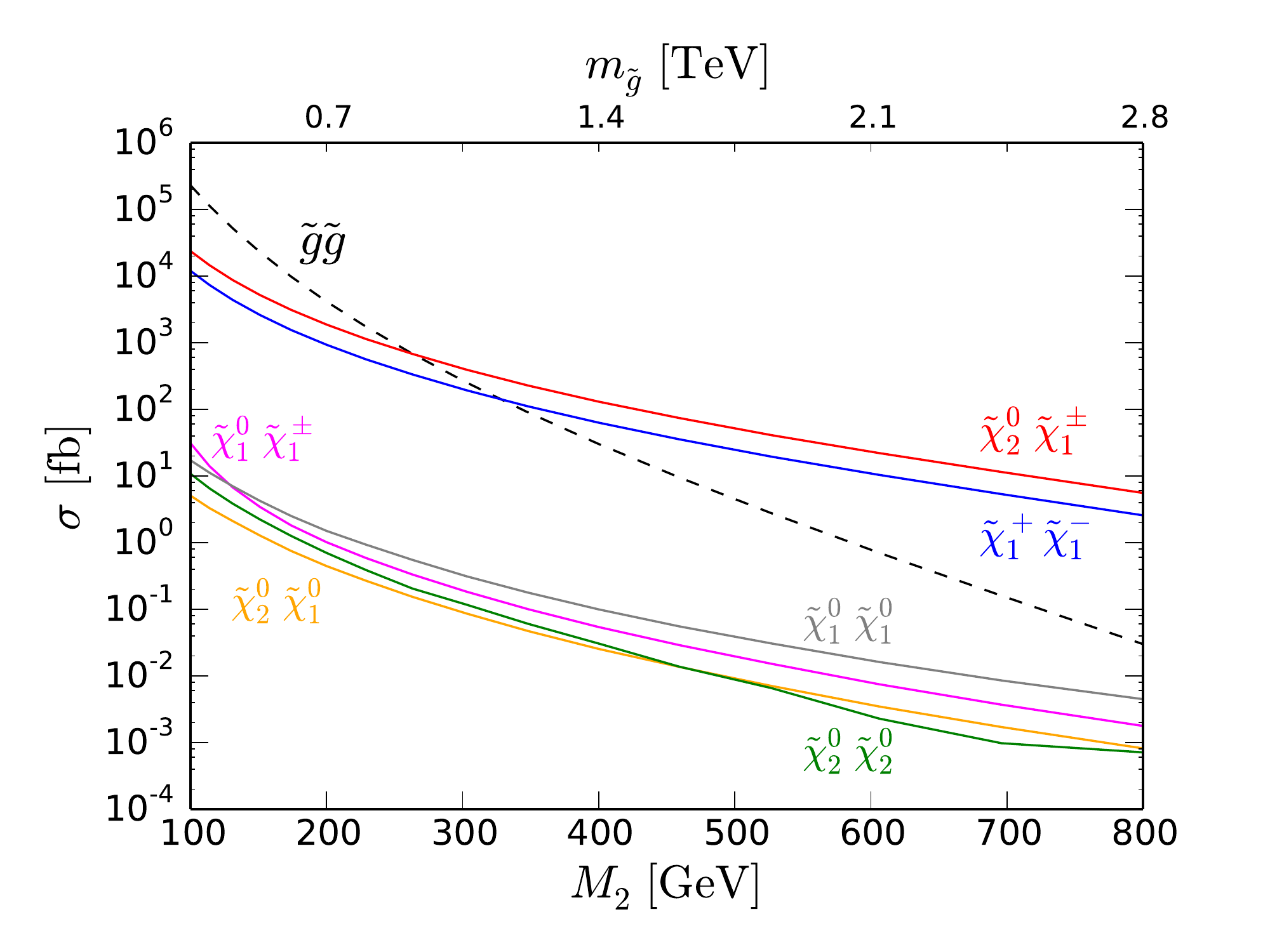}
	\caption{The NLO cross sections for various
gaugino production modes at the 14~TeV LHC as functions of $M_2$ (and $m_{\tilde g}
\simeq M_3 = 7 M_2 / 2)$. We have assumed the gaugino GUT relation,
$M_3:M_2:M_1 = 7:2:1$, at the EW scale. The other relevant parameters were fixed as $m_{\tilde q} = \mu = 3$ TeV and $\tan \beta = 10$.}
	\label{fig:xsec_all}
	\end{center}
\end{figure}

Fig.~\ref{fig:xsec_all} shows the NLO cross sections for various
gaugino production modes at the 14~TeV LHC. We have assumed the gaugino GUT relation,
$M_3:M_2:M_1 = 7:2:1$, at the EW scale and plotted the cross sections as functions of $M_2$ (and $m_{\tilde g} \simeq M_3 = 7 M_2 / 2)$.
The other relevant parameters were fixed as $m_{\tilde q} = \mu = 3$ TeV and $\tan \beta = 10$.

One can see that the $\Ntwo \Cone$ and $\tilde \chi_1^+ \tilde \chi_1^-$ production modes have substantial cross sections.
Because of the large mass hierarchy in the gaugino GUT relation, the $\gluino \gluino$ cross section drops much faster than 
the EW gaugino production cross sections as $M_2$ increases.
Due to this effect, $\Ntwo \Cone$ and $\tilde \chi_1^+ \tilde
\chi_1^-$ production dominate over $\gluino \gluino$ production for $M_2 \gsim 350$ GeV.

The EW gaugino production modes other than $\Ntwo \Cone$ and $\tilde \chi_1^+ \tilde \chi_1^-$ have cross sections which are
a few orders of magnitude smaller. This is due to the fact that these production modes contain at least one bino state or two $\tilde W^0$ states in the large $\mu$ limit,
and there exists no gaugino-gaugino-gauge boson couplings for those states.   

As can be seen from Figs.~\ref{fig:xsec} and \ref{fig:xsec_all},  the
$\Ntwo \Cone$ cross section is more than two times larger than the
$\tilde \chi_1^+ \tilde \chi_1^-$ cross section. This is mainly because $\Ntwo \Cone$ contains two distinctive modes: 
$\Ntwo \tilde \chi_1^+$ with $W^+$ exchange and $\Ntwo \tilde \chi_1^-$ with $W^-$ exchange.
It is therefore more beneficial to target the $\Ntwo \Cone$ production mode in the EW gaugino searches.

\subsection{The branching ratios}\label{sec:br}

If scalar fermions and the MSSM Higgs bosons (other than the SM-like one) are heavier than the $\Cone$ and $\Ntwo$,
these gaugino states decay predominantly into $\None$ and SM bosons, $W^\pm$, $Z$ and $h$, if the decays are kinematically allowed.
In this case, $\Cone$ exclusively decays into $W^\pm$ and $\None$ with ${\rm BR} \sim 100$\,\%.
On the other hand, $\Ntwo$ has two possible decay modes: $\Ntwo \to Z \None$ and $\Ntwo \to h \None$. 
The decay rates of these modes are determined by the $\Ntwo \None Z/h$ couplings, up to the
phase space factor and the polarisation effect.
In the limit of large $|\mu|$ and heavy MSSM Higgs bosons, the $\Ntwo \None Z/h$ couplings in the CP-conserving case are given by~\cite{Djouadi:2001fa, Bharucha:2013epa}
\beqn
|C_{\None \Ntwo Z}| &\simeq& \frac{e}{2} \frac{m_Z^2}{|\mu|^2}, \nonumber
\\
|C_{\None \Ntwo h}| &\simeq& \frac{e}{2} \frac{m_Z}{|\mu|} \Big| 2 \sin 2 \beta + \frac{M_1 + M_2}{\mu} \Big|,
\label{eq:coupling}
\eeqn
where $e$ is the electric charge ($\alpha_{em} = e^2/(4 \pi))$.

\begin{figure}[!t]
  \begin{center}
    \includegraphics[clip, scale=0.41]{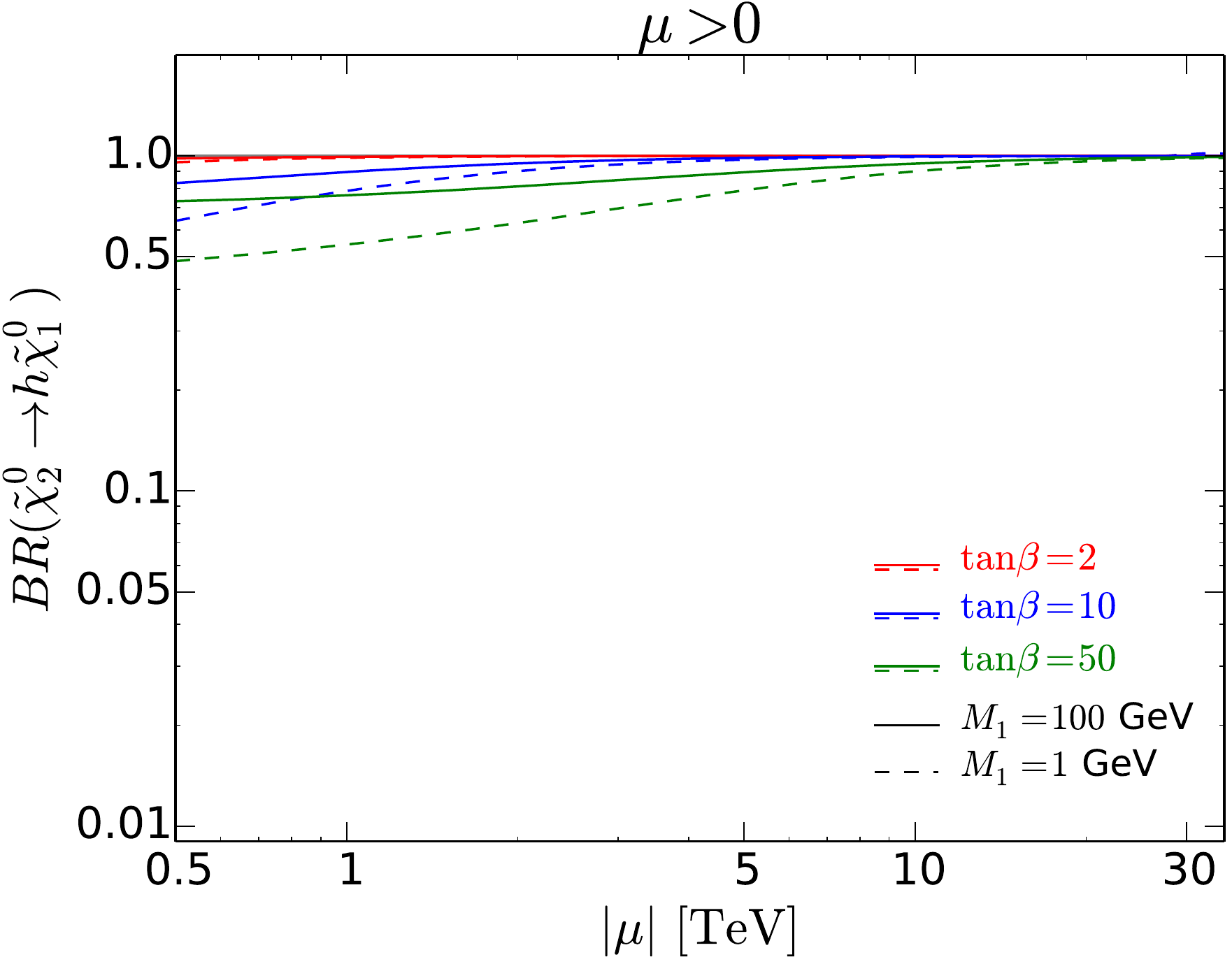}
    \hspace{1mm}
    \includegraphics[clip, scale=0.41]{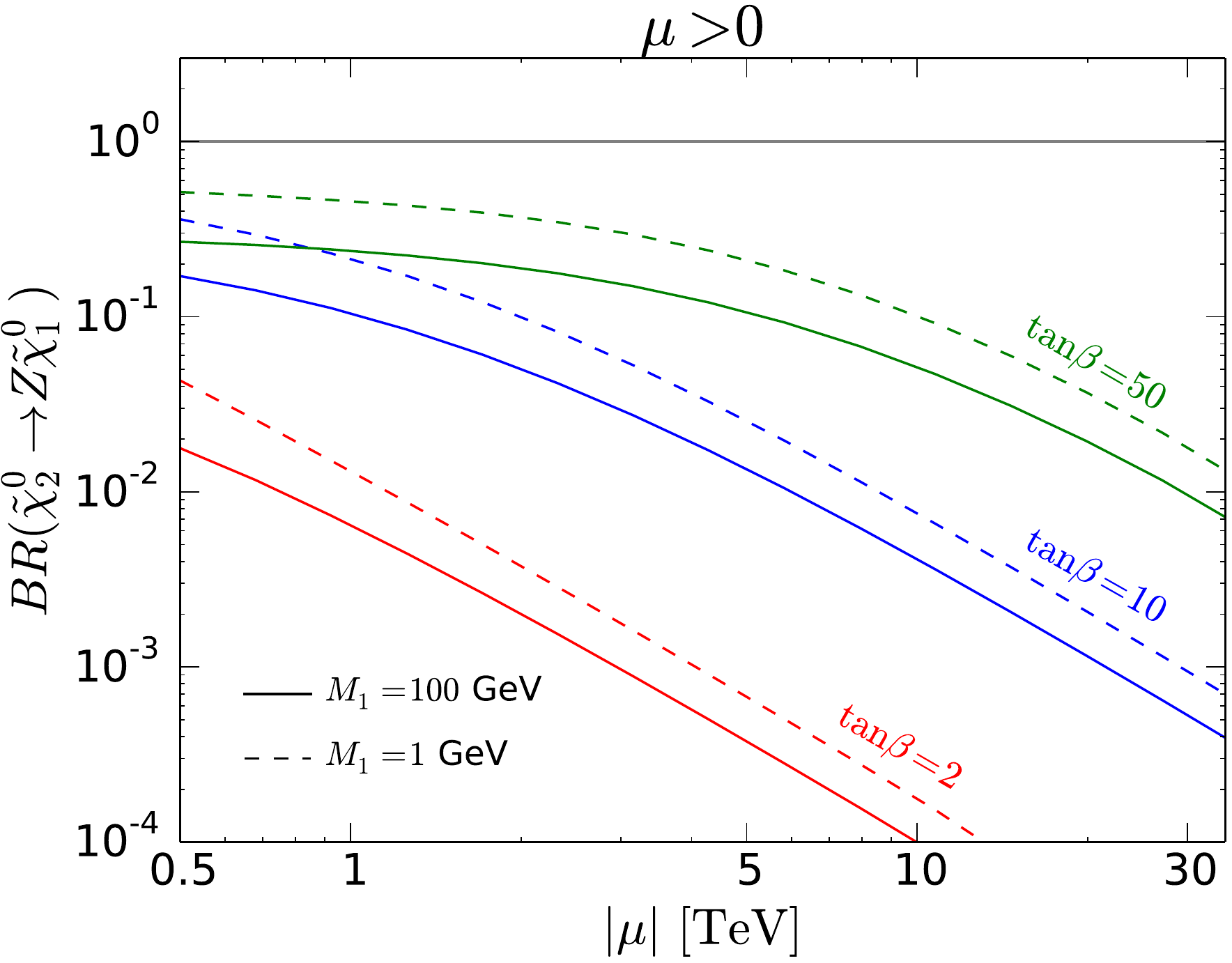}
  \end{center}
  \caption{The branching ratios of $\Ntwo \to h \None$ (left) and $\Ntwo \to Z \None$ (right) modes
as functions of $|\mu|$ in the $\mu > 0$ case. $M_2$ is taken to be
350~GeV and $m_h = 125.5$~GeV. We have fixed $M_2 = 350$~GeV but we
show variations of $\tan\beta$ and $M_1$ as $\tan\beta = 2$ (red), 10
(blue), 50 (green) and $M_1 = 100$~GeV (solid), 1~GeV (dashed).}
  \label{fig:mu-posi}
\end{figure}

Fig.~\ref{fig:mu-posi} shows the branching ratios of $\Ntwo \to h \None$ (left) and $\Ntwo \to Z \None$ (right) modes
as functions of $|\mu|$ in the $\mu > 0$ case. $M_2$ has been fixed to
$M_2 = 350$~GeV but $\tan\beta$ and $M_1$ are varied as
$\tan\beta = 2$ (red), 10 (blue), 50 (green) and $M_1 = 100$~GeV (solid), 1~GeV (dashed). 
Here and throughout the paper, we have explicitly set $m_h = 125.5$~GeV.
This condition can be always realised by tuning the stop mass, which
has no effect on the EW gaugino sector, and hence on our phenomenological analysis.
The branching ratios were calculated using {\tt SUSY-HIT}~\cite{Djouadi:2006bz}.

One can see that the $\Ntwo \to h \None$ mode is enhanced, whilst the $\Ntwo \to Z \None$ mode is suppressed
as $|\mu|$ increases.
This is due to the $\None \Ntwo Z$ coupling having the extra $m_Z/\mu$
suppression factor compared to the $\None \Ntwo h$ coupling, as seen in Eq.~(\ref{eq:coupling}).   
In the $|\mu| \gsim 500$~GeV region, the $\Ntwo \to h \None$ mode has
${\rm BR} \gsim 60\,\%$ and dominates the $\Ntwo$ decay, apart from the $\tan\beta = 50$, $M_1 = 1$~GeV case.
For moderate values of $\mu$, $0.5 \lsim \mu/{\rm TeV} \lsim 3$,
the factor $|2 \sin 2 \beta + (M_1 + M_2)/\mu|$ in $C_{\None \Ntwo h}$ is important in the competition between the $\Ntwo \to h \None$ and $\Ntwo \to Z \None$
modes 
and at $\mu \sim 1$~TeV, $\tan\beta \sim 50$, $M_1 \sim 1$~GeV,
$BR(\Ntwo \to Z \None)$ can be as large as $BR(\Ntwo \to h \None)$. 
However, in the large $|\mu|$ limit $BR(\Ntwo \to h \None)$ approaches 100\,\% independently of $\tan\beta$ and $M_1$ as long as the phase space is open.

\begin{figure}[!t]
  \begin{center}
    \vspace*{1ex}  
    \includegraphics[clip, scale=0.41]{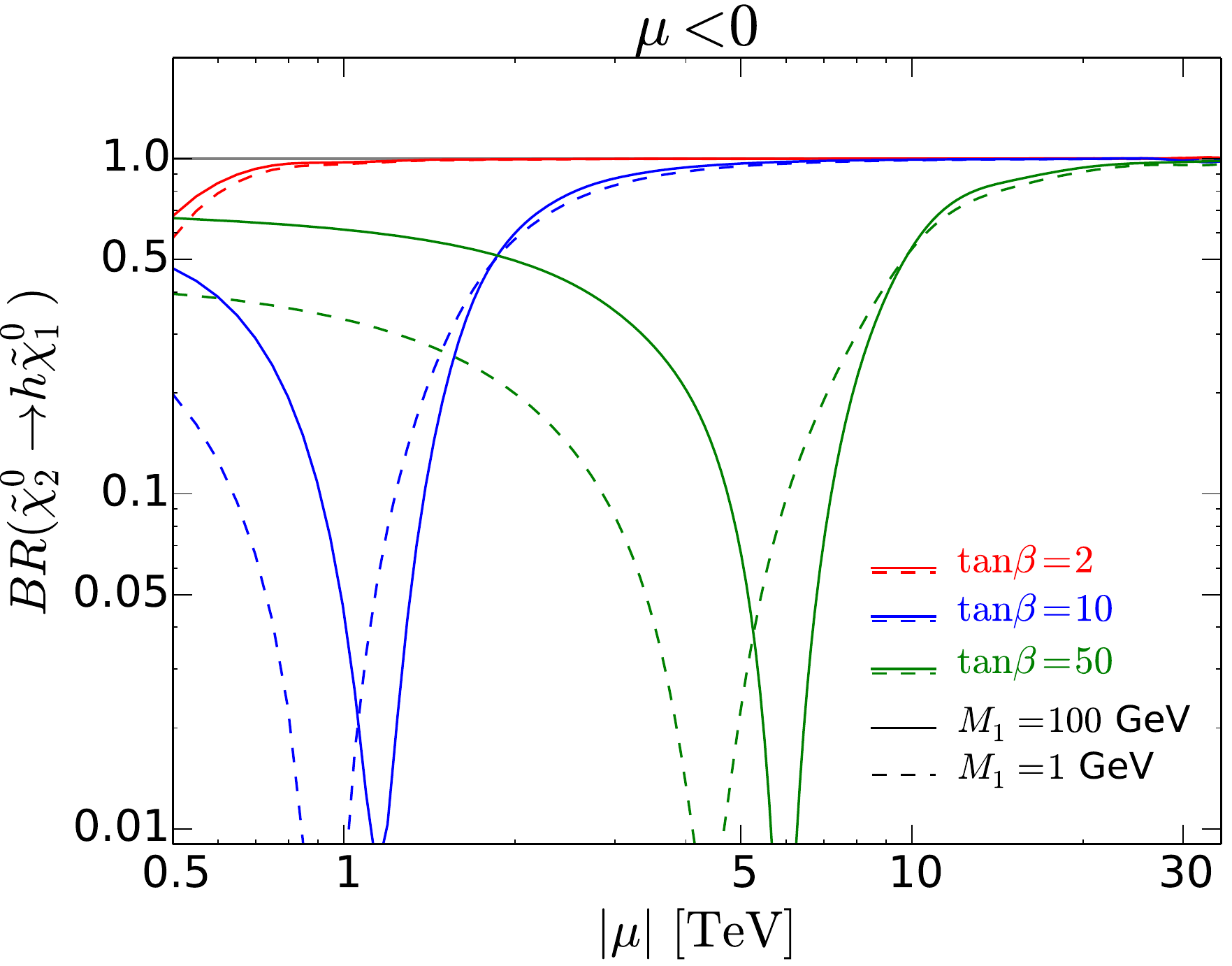}          
    \hspace{1mm}    
    \includegraphics[clip, scale=0.41]{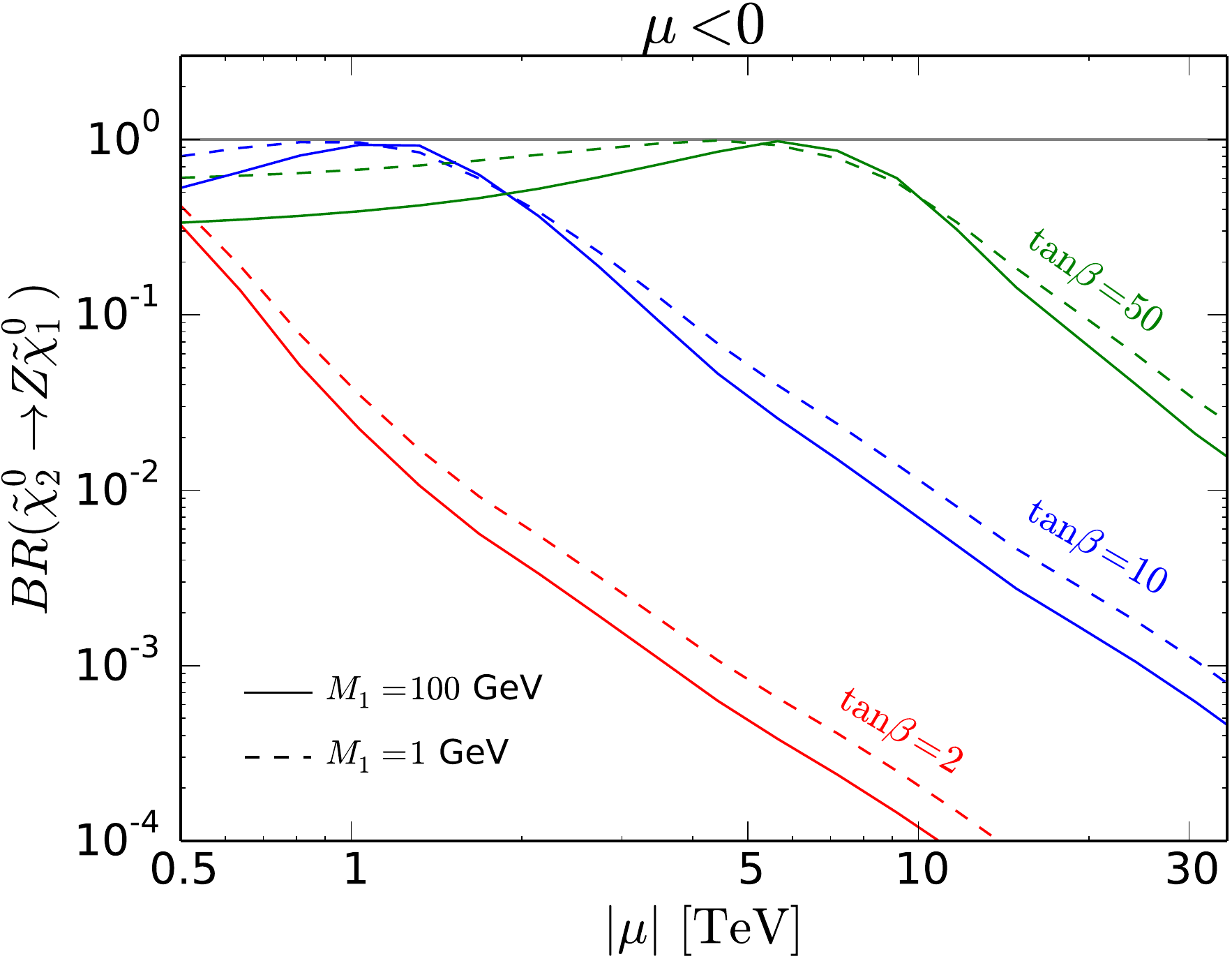}    
    \vspace{-2ex}    
  \end{center}
  \caption{Equivalent plots to Fig.~\ref{fig:mu-posi} but with $\mu <
    0$.}
  \label{fig:mu-neg}
\end{figure}

Figure~\ref{fig:mu-neg} is equivalent to Fig.~\ref{fig:mu-posi},
with $\mu$ instead set to $\mu < 0$.  One can see that $BR(\Ntwo \to h \None)$ becomes zero at a particular $|\mu|$ value depending on $\tan \beta$ and $M_1$.
This is due to the cancellation between the two terms in the $|2 \sin 2 \beta + (M_1 + M_2)/\mu|$ factor in the $\None \Ntwo h$ coupling.
As can be seen, this cancellation occurs at $\mu \sim -1$~TeV for $\tan\beta \sim 10$ and $\mu \sim - 5$~TeV for $\tan \beta \sim 50$.
As $|\mu|$ increases, $BR(\Ntwo \to h \None)$ quickly approaches 100\,\% following the cancellation.
For $|\mu| \gsim 10$~TeV, the $\Ntwo \to h \None$ mode dominates over the $\Ntwo$ decay, independently of $\tan \beta$ and $M_1$.

To summarise, we have demonstrated that in the scenarios with large $m_{\tilde q}$ and $|\mu|$, 
$\Ntwo$ and $\Cone$ become wino-like gauginos with $m_{\Ntwo} \simeq m_{\Cone} \simeq M_2$
and $\Ntwo \Cone$ has the largest cross section among the EW gaugino production modes.
We also argued that in such scenarios $\Cone$ predominantly decays into $W^\pm$ and $\None$
and the $\Ntwo \to h \None$ mode typically dominates the $\Ntwo$ decay. 
These arguments provide a strong motivation to study the $pp \to \Ntwo \Cone \to (h \None)(W^\pm \None)$ mode 
in the EW gaugino searches in the scenarios with large $m_{\tilde q}$ and $|\mu|$.

In the following sections, we study the $pp \to \Ntwo \Cone \to (h
\None)(W^\pm \None)$ channel using the $W \rightarrow \tau/\ell, \nu$ plus $h \to \tau
\tau$ and $h \to W W \to (\tau/\ell, \nu)(\tau/\ell, \nu)$ channel. 
We set $m_{\tilde q} = \mu = m_A = 3$~TeV, $\tan\beta = 10$ throughout.
This leads to $BR(\Cone \to W^\pm \None) \simeq BR(\Ntwo \to h \None) \simeq 100$\,\%. With this parameter choice, the lightest CP-even Higgs becomes SM-like
and we use the same branching ratios as those for the SM Higgs boson.  
Although the above parameter set is motivated by the heavy scalar
scenario, our analysis can easily be recast onto other SUSY
scenarios. Changing the above parameters may modify the $pp \to \Ntwo \Cone$ cross section and
the $\Ntwo \to h \None$ branching ratio significantly but does not alter the signal efficiencies for the signal regions defined in the next section.
The discovery reach and exclusion limit for a different set of parameters can therefore be obtained by rescaling the cross section and branching ratio accordingly.
Moreover, the calculated signal efficiencies can also be used for a larger
class of models as far as the $\tilde N \tilde C^{\pm} \to (h \chi)
(W^\pm \chi)$ topology is concerned, as mentioned in
subsection~\ref{sec:setup}. Neglecting a finite width effect and spin correlations, the signal efficiencies will be very similar between 
$\Ntwo \Cone \to (h \None) (W^\pm \None)$ and  $\tilde N \tilde
C^{\pm} \to (h \chi) (W^\pm \chi)$ at $(m_{\Ntwo} = m_{\Cone},
m_{\None}) = (m_{\tilde N} = m_{\tilde C}, m_{\chi})$. We explain this point in more detail in Appendix~\ref{app:sr} and
provide the necessary information to perform such a re-analysis.

\section{Simulation and analysis}\label{sec:analysis}

\subsection{Monte Carlo simulation}
The SUSY $pp \to \Ntwo \Cone \to (h
\None)(W^\pm \None)$ signals were generated using the \HWPP general-purpose event
generator~\cite{Bahr:2008pv, Arnold:2012fq, Bellm:2013lba} via SUSY
Les Houches Accord files used as input for the parameter points,
according to the assumptions outlined in the previous section. The
signal cross sections were scaled to the next-to-leading order cross
sections using results obtained from \texttt{Prospino}
2.1. The $hV$,
$t\bar{t}$, $t\bar{t}h$ and $WZ$ backgrounds were also generated internally
in \textsf{HERWIG++} at leading order. The $Z$+jets and $W$+jets backgrounds were generated using
the parton-level matrix element generator \texttt{AlpGen} and merged with the
\textsf{HERWIG++} parton shower using the MLM
method~\cite{Mangano:2001xp,Mangano:2004lu,Alwall:2007fs}. The
generator-level cuts on the $V$+jets backgrounds were taken to be
$p_{Tj,\mathrm{min}} = 15$~GeV, $\eta _{j,\mathrm{max}} = 3.0$, $\Delta R _{j,\mathrm{min}} =
0.2$ with $m_{\ell\ell} \in (15, 160)$~GeV (or $m_{\tau,\tau}$) for $V=Z$. For the $Z$+jets case we considered matrix elements with one
extra parton merged to the shower, whereas for the $W$+jets case we
considered matrix elements with two partons merged to the shower. 

For the signal we allowed the $W$ to decay to all lepton flavours,
including taus. Likewise, for the backgrounds we consider all of the
leptonic decays of the $W$ and $Z$, to muons, electrons or taus. We
consider the Monte Carlo samples of the $Z$ and $W$ backgrounds going
to electrons or muons separately from those going to taus, as they
would have different amounts of missing energy, leptons and jets. The Higgs boson was allowed to decay to
$\tau^+\tau^-$ or $W^+W^-$ with subsequent decay of the $W$ bosons to
$e \nu_e$, $\mu \nu_\mu$ and $\tau \nu_\tau$.  

In all cases of signal and background the full parton shower, hadronization and the underlying
event~\cite{Bahr:2008dy} were included.\footnote{We do not include a
  description of pile-up events. These should be considered in detail
  in a full experimental simulation.} All the runs have been generated using the MSTW2008nlo
68\% PDF set.  We note that we do not consider pure QCD-initiated
backgrounds since these are expected to be negligible in the
high-missing transverse momentum regime, particularly in conjunction with the
existence of isolated leptons or $\tau$s. 

We define a SUSY benchmark point \textbf{C350-100}, with
parameters
\beq
M_2 = m_{\Cone} = m_{\Ntwo} = 350~{\rm GeV},~~~ M_1 = m_{\None} = 100~{\rm GeV}.
\eeq
This point will be used as an example to demonstrate the effect of cuts
and provide a typical point to aid the development of the strategy for
discriminating the signal against the various backgrounds. 
\subsection{Tau identification}\label{sec:tauid}
\subsubsection{Tau lepton decay modes}
The study of final states containing hadronically decaying $\tau$
leptons is an important and growing part of the LHC's physics
program. The $\tau$ lepton has a multitude of
decay modes, which we may split these into two categories: `leptonic', if
the visible decay products contain a single lepton, and `hadronic', if
there are one or three charged hadrons present. We label the corresponding
modes $\tau_\ell$ and $\tau_h$ respectively. The hadronic modes are also
categorised as `1-prong' and `3-prong', according to the number of
charged particles involved in the decay.\footnote{
If one of the taus undergoes a 3-prong decay, one may improve the analysis significantly using 
the information of the secondary vertex of the 3-prong tau decay~\cite{Gripaios:2012th}.
This requires a dedicated study and we do not use the secondary vertex information in this paper. 
}  The label `$\ell$' here and elsewhere implies an
electron or a muon. The branching ratios for these modes
are:\footnote{These do not add up to 100\%, since we are only considering the dominant 1-prong and 3-prong
  decay modes.}
\begin{itemize}
\item leptonic: BR($\tau \rightarrow \tau_\ell$)~$\sim 0.35$. 
\item hadronic: BR($\tau \rightarrow  \tau_h$)~$\sim 0.625$. 
\end{itemize}
These imply, for a Higgs boson decay to $\tau^+\tau^-$: 
\begin{itemize}
\item BR($h \rightarrow \tau_h \tau_h $)~$\sim 0.39 \times
  BR(h\rightarrow \tau^+\tau^-)$. 
\item BR($h \rightarrow  \tau_h \tau_\ell$)~$\sim 0.44 \times
  BR(h\rightarrow \tau^+\tau^-)$. 
\item BR($h \rightarrow  \tau_\ell \tau_\ell$)~$\sim 0.12 \times
  BR(h\rightarrow \tau^+\tau^-)$. 
\end{itemize}
\subsubsection{Hadronic tau identification algorithm}
Both ATLAS~\cite{TheATLAScollaboration:2013wha} and CMS~\cite{CMS:tva}
employ reconstruction and identification
algorithms, used to identify hadronically decaying $\tau$ leptons and
reject various backgrounds. Here, we do not attempt to reproduce either of the
ATLAS or CMS algorithms exactly, but instead use elements from both resulting
in an algorithm that we expect performs in an equivalent way. We also borrow
elements from~\cite{Katz:2010iq}, which examines di-$\tau$ tagging in the boosted
regime.\footnote{For di-$\tau$ tagging in Higgs searches, see also Ref.~\cite{Englert:2011iz}.} The resulting algorithm is expected to provide conservative
hadronic $\tau$-tagging results, and could be improved substantially
via the use of boosted decision trees (BDT) or other advanced
multivariate methods. Since we will not employ simulation of detector
effects in the present analysis, we focus on a simple cut-based
algorithm for simplicity.

The first part of the basic algorithm for hadronic $\tau$
identification proceeds as follows:
\begin{itemize}
\item Reconstruct jets with $R=0.5$ using the Cambridge/Aachen jet
  algorithm as implemented in \texttt{FastJet}~\cite{Cacciari:2011ma}. An individual jet is then
  investigated for constituent hadronic tracks.\footnote{Usage of the
    word ``track'' here and elsewhere in this article implies
    ``charged particle''.} 
\item Consider a track to be a `seed' if it is the hardest track
  in the jet, has $p_T > 5$~GeV and is within $\Delta R = 0.1$ of the
  jet axis. 
\item If such a track is found, one defines inner and outer cones
  around it. We use $R_{in} = 0.2$ and $R_{out} = 0.4$ respectively. 
\item Require \textit{no} photons with $p_T > 2$~GeV and \textit{no} charged
  tracks with $p_T > 1$~GeV to lie within the defined annulus between
  $R_{in}$ and $R_{out}$. 
\end{itemize}
The basic part of the algorithm itself does not provide satisfactory
rejection against the QCD jet background to hadronically decaying $\tau$
leptons. If a jet satisfies all the above criteria, then the following
variables are constructed:
\begin{itemize}
\item $\Delta R_{\mathrm{max}}$: the distance to the track furthest
  away from the jet axis. 
\item $f_{\mathrm{core}}$: the fraction of the total jet energy
  contained in the centre-most cone defined by $\Delta R < 0.1$. 
\end{itemize}
These variables provide strong discriminating power
against QCD jets~\cite{TheATLAScollaboration:2013wha,
  ATLAS:2011oka}.\footnote{More variables have been employed by the
  experimental collaborations, but we found that the two that we
  consider are sufficient at this level of simulation.} To perform the
rejection of QCD jets, here we apply the following cuts: 
\begin{itemize}
\item $\Delta R_{\mathrm{max}} < 0.05$.
\item $f_{\mathrm{core}} > 0.95$.
\end{itemize}

\begin{figure}[t]
  \begin{center}
    \vspace*{1ex}
    \includegraphics[scale=0.45]{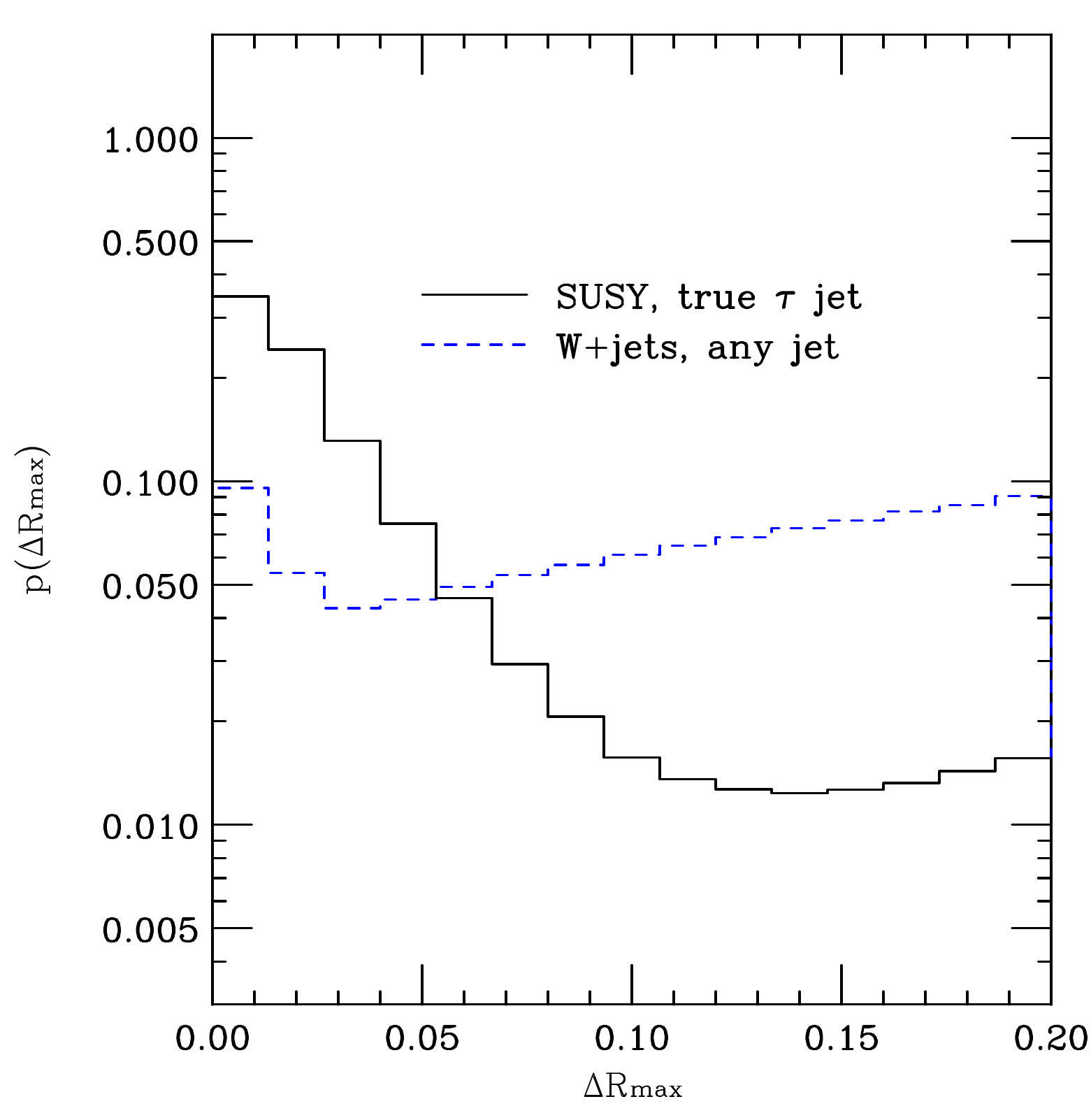}
    \includegraphics[scale=0.45]{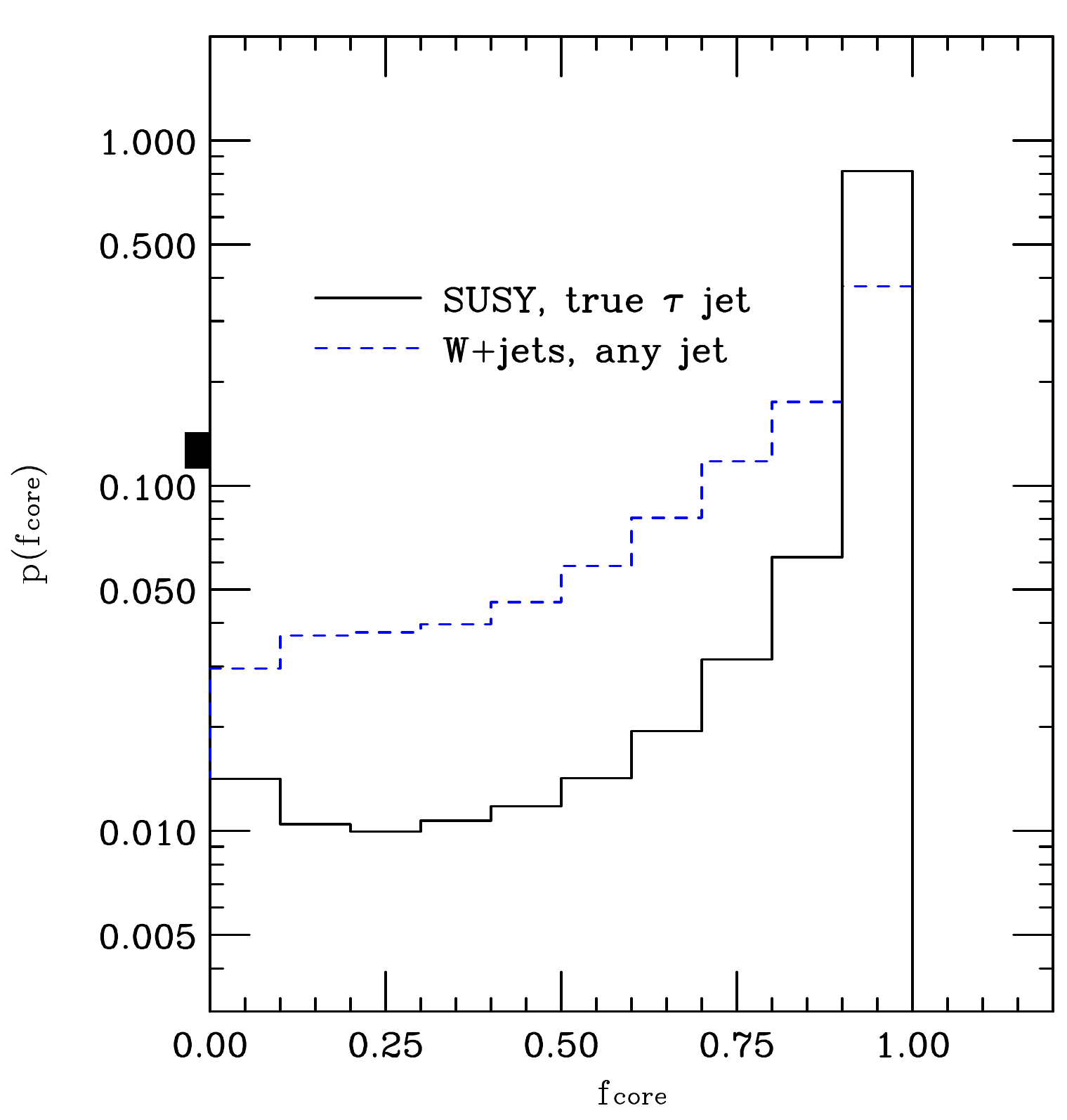}
    \vspace{-2ex}
  \end{center}
  \caption{Distributions of the variables used for discrimination of the jets
    originating from $\tau$ leptons and those from QCD, for the SUSY
    benchmark point \textbf{C350-100} and the $W$+jets background ($W
    \rightarrow e\nu_e/\mu \nu_\mu$).}
  \label{fig:variables}
\end{figure} 
\begin{figure}[t]
  \begin{center}
    \vspace*{1ex}
    \includegraphics[scale=0.50]{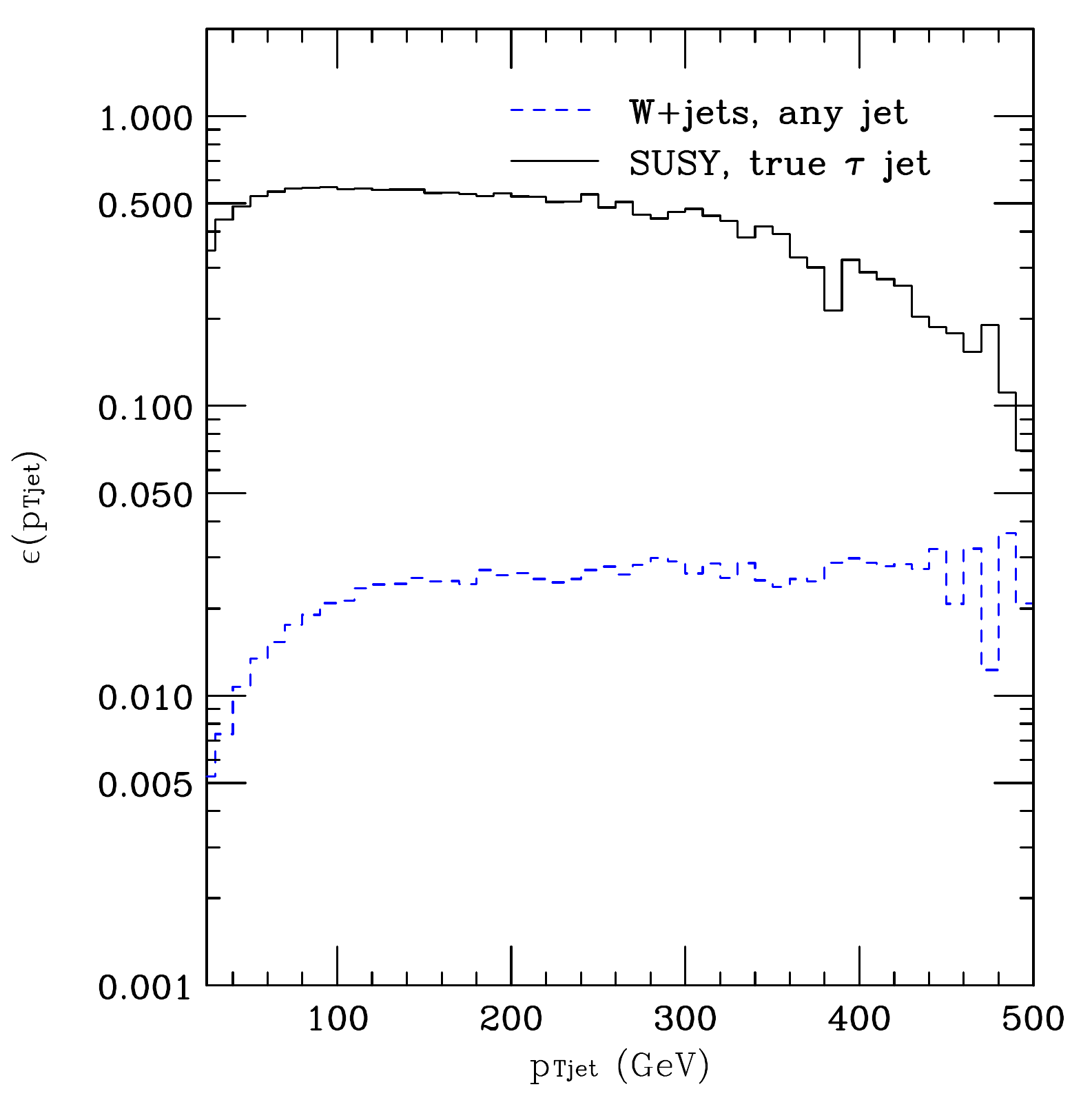}
    \vspace{-2ex}
  \end{center}
  \caption{The efficiency of tagging a jet as a $\tau$-jet for the SUSY
    benchmark point \textbf{C350-100} and the $W$+jets background
    (with $W
    \rightarrow e\nu_e/\mu \nu_\mu$). For \textbf{C350-100}, the efficiency was defined for the identification of `true'
$\tau$ jets, defined to be those closest to the visible $\tau$ decay products
taken from the Monte Carlo truth. For the $W$+jets, the
efficiency was defined with respect to any jet. Jets of $p_T > 20$~GeV
and $| \eta | < 2.5$ are considered in both cases.}
  \label{fig:efficiency} 
\end{figure} 

In Fig.~\ref{fig:variables} we show the variables $\Delta
R_{\mathrm{max}}$ and
$f_{\mathrm{core}}$, constructed for hadronic
jets for a signal benchmark point \textbf{C350-100} and the $W$+jets
background.
Only jets with $p_T > 20$~GeV and $| \eta | < 2.5$ were considered. In
Fig.~\ref{fig:efficiency} we show the efficiency of $\tau$ identification versus the transverse momentum of
the jet in question, $p_{T,\mathrm{jet}}$, obtained by the procedure outlined in this
section. For the signal, the efficiency was defined for the identification of `true'
$\tau$ jets, defined to be those closest to the visible $\tau$ decay products
taken from the Monte Carlo truth. For the $W$+jets background, the
efficiency was defined with respect to any jet. The efficiency for the SUSY
benchmark point \textbf{C350-100} varies from around 50\% in the
$p_{T,\mathrm{jet}}$ region of $20-300$~GeV and
then drops down to $\sim 20\%$ at around $p_{T,\mathrm{jet}} \sim 400$~GeV. For the $W$+jets background
the efficiency starts off at $\sim 1\%$ at $p_{T,\mathrm{jet}}\sim20$~GeV and then
rises to an efficiency of $2-3\%$, more or less constant up to
$p_{T,\mathrm{jet}}\sim 500$~GeV.

\subsection{Analysis}
Since the signal events contain hard jets or isolated hard leptons,
they are expected to pass the experimental triggers with
high efficiency, and hence we do not consider the effect of triggering
here. We define the first level of the analysis for discriminating the
signal against the various backgrounds as follows:
\begin{enumerate}
\item Particles of $p_T > 0.5$~GeV and $|\eta| < 5.0$ are considered. 
\item If isolated leptons with $p_T > 20$~GeV are found, they are placed in a separate list, and removed from the list of
  particles. An isolated lepton is defined as either: having $\sum _i p_{T,i}$
  less than 20\% of its transverse momentum around a cone of $\Delta R
  = 0.4$ around it, or as a lepton that contains no photons with $p_T > 2$~GeV and no tracks with $p_T > 1$~GeV in the annulus $\Delta R = (0.2, 0.4)$ around it.\footnote{We apply two different criteria to take into account the possibility of radiation from the core lepton.}
\item Jet finding is performed on the list of remaining particles, using
  \texttt{FastJet} and the
  Cambridge/Aachen jet algorithm, with parameter $R=0.5$. Jets of $p_T
  > 20$~GeV are accepted. 
\item Tagging of $\tau$-jets is performed as described in
  Section~\ref{sec:tauid}. 
\item Only events with a total number of isolated leptons, $n_{\ell,\mathrm{iso}}$, and
  $\tau$-tagged jets, $n_{\tau,\mathrm{tag}}$, equal to 3 are
  accepted: i.e. we require $n_{\tau,\mathrm{tag}} + n_{\ell,\mathrm{iso}} = 3$. A hypothesis is then performed to match the topology of the
  SUSY events. The hypotheses vary according to the number
  of isolated leptons and $\tau$-tagged jets and are listed in detail in
  Table~\ref{tb:hypotheses}. 
\item Several variables are calculated and are passed through to
  the second level of analysis. 
\end{enumerate}

\begin{table}[!htb]
  \begin{center}
    \begin{tabularx}{\linewidth}{llXX}
      \toprule
       $n_{\tau,\mathrm{tag}}$ & $n_{\ell,\mathrm{iso}}$ & real signal
       channels & hypothesis\\ \midrule
       3 & 0 & ($h \rightarrow \tau_h \tau_h$, $W\rightarrow \tau \nu$) & assign hardest two to $h$. \\
       2 & 1 & ($h \rightarrow \tau_h \tau_\ell$, $W\rightarrow \tau_h
       \nu_\tau$), \newline ($h\rightarrow \tau_h \tau_h$, $W\rightarrow \ell \nu_\ell$)
       & assign hardest two to $h$. \\
       1 & 2 & ($h\rightarrow \tau_h \tau_\ell$, $W\rightarrow \ell
       \nu_\ell$), \newline $(h\rightarrow \tau_\ell \tau_\ell$, $W\rightarrow
       \tau_h \nu_\tau$) & if leptons are same sign, assign
         highest-$p_T$ 
       to $h$ along with the $\tau$-tagged jet. Otherwise: assign any
       two highest-$p_T$ to $h$. \\
       0 & 3 & ($h \rightarrow \tau_\ell \tau_\ell$, $W\rightarrow
       \ell \nu_\ell$) &  If all leptons are the same sign, reject the
       event. Otherwise: pair two highest-$p_T$ of opposite sign as
       the $h$.\\ \bottomrule
    \end{tabularx}
  \end{center}
  \caption{The hypotheses applied for the reconstruction of the
    Supersymmetric topology as described in the main text. The
    different hypotheses are given according to the number of
    $\tau$-tagged jets, $n_{\tau,\mathrm{tag}}$, and the number of isolated
    leptons $n_{\ell,\mathrm{iso}}$. In the final stage of the
    analysis, the $n_{\tau,\mathrm{tag}} = 3$ was found to reduce
    significance and was not considered.}
\label{tb:hypotheses}
\end{table}

Steps 1-5 are what we define as the `basic' analysis. The variables
calculated in step 6 and used for further discrimination in the second-level analysis are: the
transverse momentum of the di-$\tau$-tagged system, $p_{T,\tau\tau}$,
the distance between the $\tau$-tagged jets, $\Delta R _{\tau,\tau}$,
the distance between the di-$\tau$-tagged system and the lepton,
$\Delta R _{\tau\tau, \ell}$, the missing transverse energy, $\slashed{p}_T$ and the variable
$M_{\mathrm{min}}$, which is sharply peaked at low values for the
$WZ$ background and broadly falls off for the signal, defined in Appendix~\ref{app:mmin}. The variables
are outlined in Table~\ref{tb:variables}. There we provide an
\textit{example} set of cuts, applied to the SUSY benchmark point \textbf{C350-100}, found to give a
significance of $\sim 2.5\sigma$ for an integrated luminosity of 100~fb$^{-1}$ at 14~TeV. For completeness, we
show in Table~\ref{tb:benchanal} the resulting cross sections after
applying the analysis on the SUSY benchmark point and the different
backgrounds for this example. In the final stage of the analysis the
$n_{\tau,\mathrm{tag}} = 3$ channel was excluded, since it was found to reduce
significance by allowing more background. Note that this set of cuts
will constitute `signal region 1' of our full analysis. 

Details of how the initial cross sections for the signal and background are calculated are given
in Appendix~\ref{app:crossx}. We note that in the case of the $Z$+jets
and $W$+jets samples, we obtained $N_\mathrm{cuts} = 0$ events after
all cuts.\footnote{For the $Z(\rightarrow \tau^+ \tau^-)$+jets, this
  depends on how large the missing transverse momentum cut imposed is.} To provide an estimate of the cross section, we assume that
the Poisson distribution has mean number of events $\lambda = 3$ and use this
as an upper bound to estimate the resulting cross sections. The
probability of having a Poisson-distributed sample with mean $\lambda > 3$,
given that zero events have been observed, is $\simeq 0.05$. It is useful to mention at this point that we do not apply a $K$-factor to the
$Z$+jets or $W$+jets cross sections. The induced uncertainty due to
this omission can be absorbed in the systematic uncertainty due to 
lack or low number of events in the final Monte Carlo samples. Nevertheless, since conservative estimates for these
backgrounds have been assumed, $K$-factors of $\sim 2$ would not have a
significant impact to the main conclusions of our analysis. 
\begin{table}[!htb]
  \begin{center}
    \begin{tabularx}{\linewidth}{lXX}
      \toprule
       variable & definition & benchmark point cut ($\equiv$ signal
       region 1)\\ \midrule
       $\slashed{p}_T$ & missing transverse momentum & $>95$~GeV \\
       $M_{\mathrm{min}}$ & Appendix~\ref{app:mmin} & $>235$~GeV\\
       $p_{T,\tau\tau}$ &  di-$\tau$-tagged jet $p_T$ & $>20$~GeV\\ 
       $\Delta R_{\tau,\tau}$ &  distance between $\tau$-tagged jets & $\in (0.1, 2.9)$\\ 
       $\Delta R_{\tau\tau,\ell}$ &  distance between di-$\tau$-tagged
       jet system and lepton  & $\in (0.1, 2.6)$\\ \bottomrule
    \end{tabularx}
  \end{center}
  \caption{The variables used for further discrimination after the
    basic part of the analysis is applied to the signal and
    backgrounds.}
\label{tb:variables}
\end{table}

\begin{table}[!htb]
  \begin{center}
    \begin{tabularx}{\linewidth}{XXXX}
      \toprule
       sample & $\sigma_\mathrm{initial}$~(fb)
       &$\sigma_\mathrm{basic}$~(fb) & $\sigma_\mathrm{cuts}$~(fb)  \\ \midrule
      SUSY \textbf{C350-100} & $5.7$ & 0.658 & 0.152\\
      $WZ$ & $767$ & 85.734 & 0.079 \\
      $W (\rightarrow \ell\nu_\ell) $+jets& $\sim 600\times10^3$ & 61.974 &$ \lesssim 0.055$ \\
      $W (\rightarrow \tau\nu_\tau) $+jets& $\sim 300\times10^3$ & 7.591  &$ \lesssim 0.052$ \\
      $hV$ & $443$ & 5.071 & 0.037\\
      $t\bar{t}h$& $3.4$ & 0.147 & 0.008 \\
      $t\bar{t}$ & $8600$ & 14.876 & 0.005\\
      $Z (\rightarrow \ell \ell)$+jets& $\sim 600\times10^3$ & 1659 & $\lesssim 0.029$ \\
      $Z (\rightarrow \tau \tau)$+jets& $\sim 300\times10^3$ & 52.762 & 0.047 \\\bottomrule
    \end{tabularx}
  \end{center}
  \caption{The effect of the cuts on the SUSY benchmark point
    \textbf{C350-100} and the relevant backgrounds. The initial cross section calculations are presented
    in Appendix~\ref{app:crossx}.}
\label{tb:benchanal}
\end{table}

\subsection{Signal regions}\label{sec:signalreg}
To perform a scan of the supersymmetric parameter space, we define signal regions, with cuts that aim to bring out the different
qualities of the defined variables. These signal regions are shown in Table~\ref{tb:signalreg}, for the variables defined in
Table~\ref{tb:variables}. All the signal regions exclude the
$n_{\tau,\mathrm{tag}} = 3$ channel, since it was found to reduce
significance. 

\begin{table}[!htb]
 \begin{center}
  \begin{tabularx}{\linewidth}{lXXXXXXX}
      \toprule
       variable & SR1 & SR2 & SR3 & SR4 & SR5 & SR6 & SR7 \\ \midrule
       $\slashed{p}_T$ & $95$~GeV & $120$~GeV & 
       $100$~GeV & $90$~GeV& $90$~GeV & $150$~GeV & $90$~GeV \\
      $M_{\mathrm{min}}$ & $235$~GeV & $270$~GeV & $220$~GeV  &
      $220$~GeV & $300$~GeV& $240$~GeV &$200$~GeV \\
       $p_{T,\tau\tau}$ & $20$~GeV & $80$~GeV & $20$~GeV  & $50$~GeV &$20$~GeV & $20$~GeV& $20$~GeV\\ 
      $\Delta R_{\tau,\tau}$ &  $ (0.1, 2.9)$ & $ (0.1, 2.9)$
      &$ (0.1, 2.9)$  &$ (0.1, 2.9)$  & $ (0.1, 2.9)$ & $ (0.1, 2.9)$ &$ (0.1, 2.9)$  \\ 
       $\Delta R_{\tau\tau,\ell}$ & $ (0.1, 2.6)$ & $ (0.1,2.5)$
       & $ (0.1, 2.6)$  &  $ (0.1, 2.6)$ &  $ (0.1, 2.6)$ & $ (0.1, 2.6)$  &  $ (0.1, 2.6)$ \\ \bottomrule
    \end{tabularx}
 \end{center}
  \caption{The cuts for the different signal regions (SR) used in the analysis.}
\label{tb:signalreg}
\end{table}

\section{Results}\label{sec:results}
\begin{figure}[!htb]
  \begin{center}
    \vspace*{1ex}
    \includegraphics[scale=0.60]{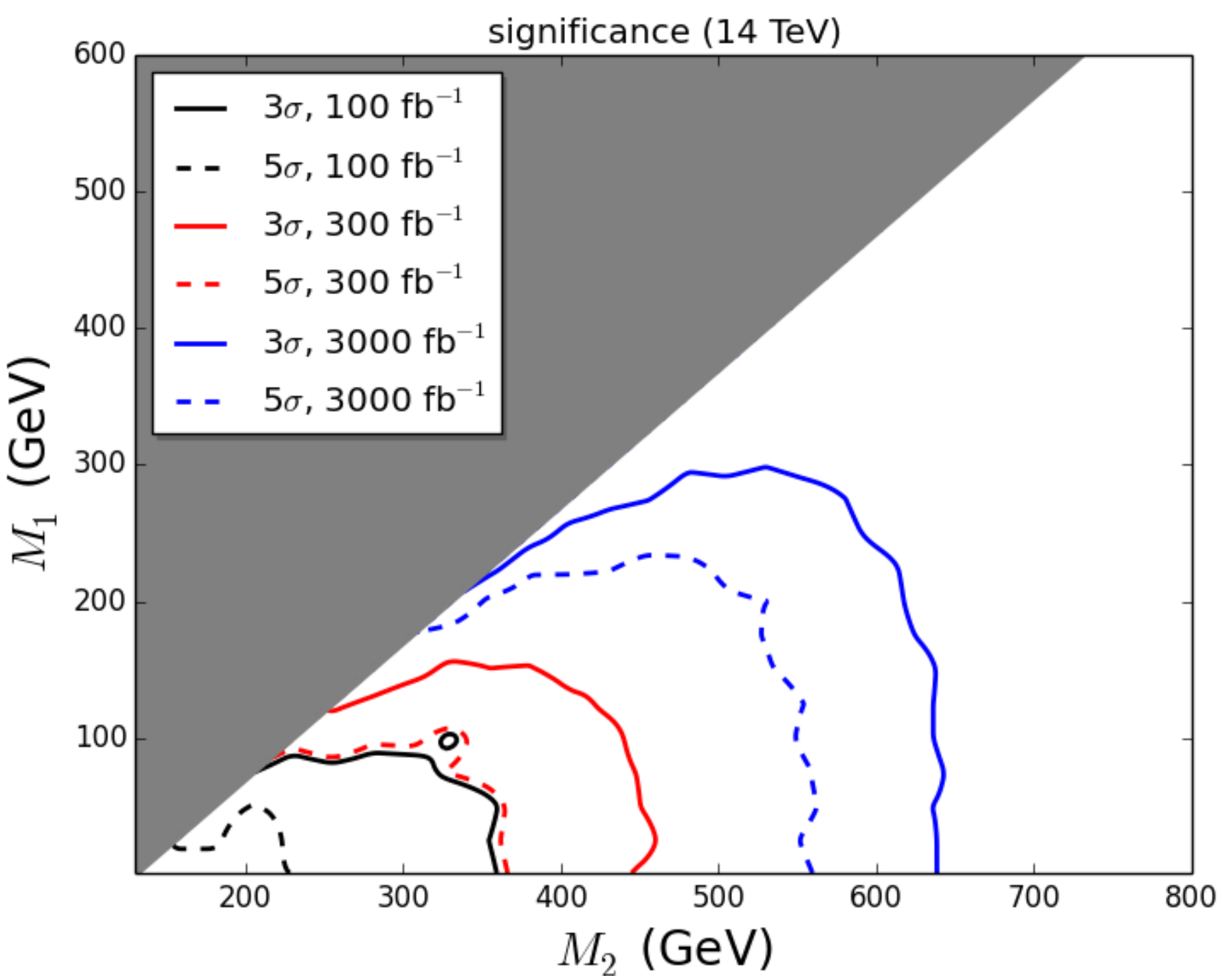}
  \end{center}
  \caption{The significance envelope on the $M_2$-$M_1$ plane obtained for the
    signal regions defined in Table~\ref{tb:signalreg} at integrated
    luminosities of 100~fb$^{-1}$ (black) or 300~fb$^{-1}$ (red). The solid curves show the $3\sigma$
   evidence region, whereas the dashed curves show the $5\sigma$ discovery region.}
  \label{fig:sig} 
\end{figure}

\begin{figure}[!htb]
  \begin{center}
    \vspace*{1ex}
    \includegraphics[scale=0.60]{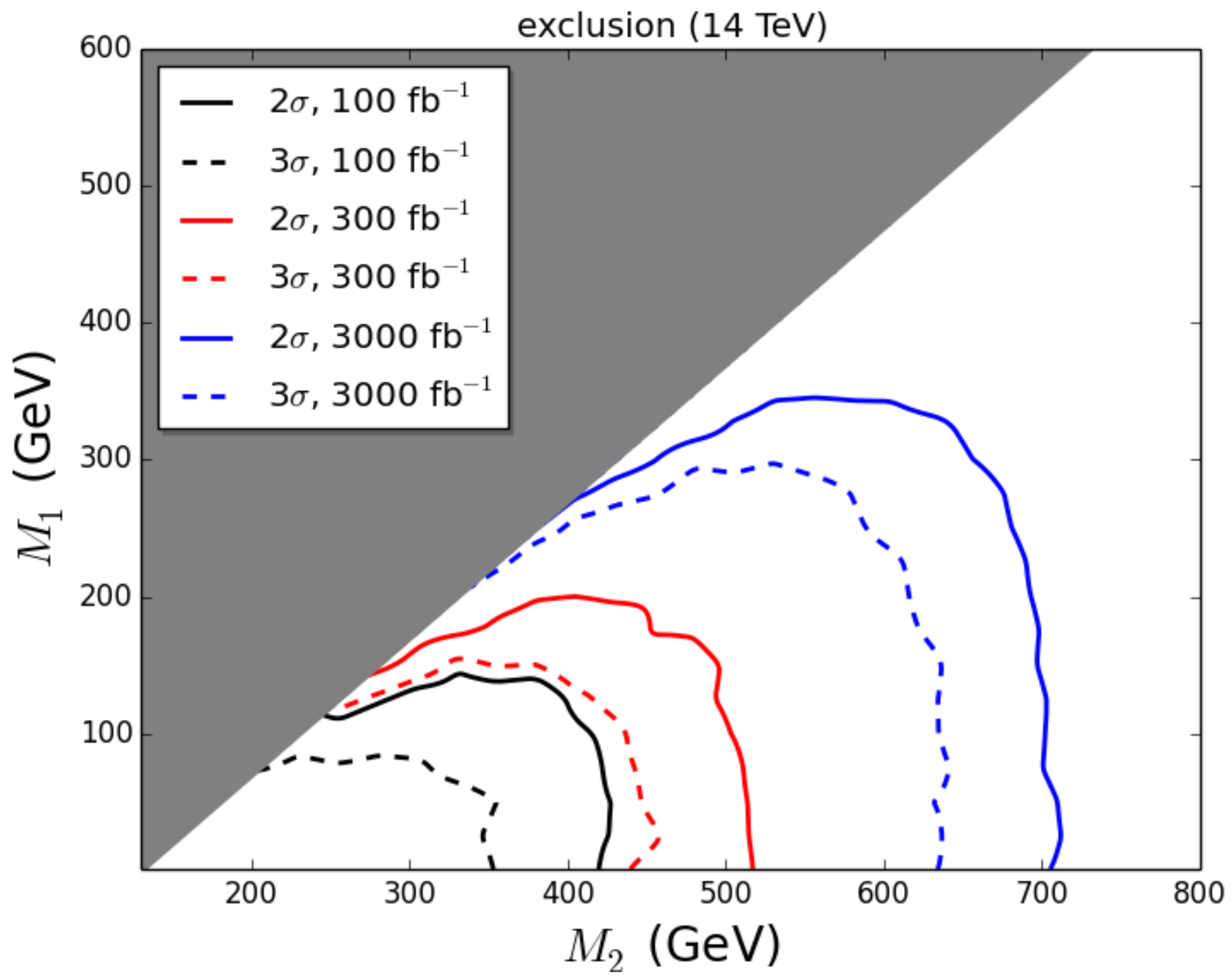}
  \end{center}
  \caption{The exclusion envelope on the $M_2$-$M_1$ plane obtained for the
    signal regions defined in Table~\ref{tb:signalreg} at integrated
    luminosities of 100~fb$^{-1}$ (black) or 300~fb$^{-1}$ (red). The solid curves show the $2\sigma$
   exclusion boundary, whereas the dashed curves show the $3\sigma$ boundary.}
  \label{fig:exc}
\end{figure}

We performed the analysis on the $M_2$-$M_1$ plane, according to the
cuts defined in the signal regions in Table~\ref{tb:signalreg} at
integrated luminosities of 100~fb$^{-1}$, 300~fb$^{-1}$ and 3000~fb$^{-1}$. We
show the resulting envelope of significances in Fig.~\ref{fig:sig}, where the solid
curves show the $3\sigma$ evidence region, whereas the dashed curves
show the $5\sigma$ discovery region. We also show in
Fig.~\ref{fig:exc}, the expected exclusion region at 2$\sigma$ (solid) and
3$\sigma$ (dashed). For completeness, we show the corresponding
overlapping signal regions in Appendix~\ref{app:sr}. There, we also
provide the total cross sections for the backgrounds after cuts given by the different signal
regions. These can be used to infer constraints in
explicit SUSY models that contain the specific decay chain we are considering. 

The analysis can yield a low number of events for both signal and background, of $\mathcal{O}(10)$, and for the calculation of
significance we used the Poisson distribution to calculate the
$p$-values. These were subsequently converted to the corresponding Gaussian
standard deviations. Details of the procedure are provided in
Appendix~\ref{app:statistics}, with supplementary material in Appendix~\ref{app:poisson}.

Although the authors of Ref.~\cite{Ghosh:2012mc} have not performed an equivalent
parameter-space scan over $M_1$-$M_2$, and the details of the chosen parameters
differ from the ones presented in this article, it is still interesting to compare with the potential of the final state in
which the channel $\Ntwo
\Cone \to (h \None) (W^\pm \None)$ involves leptonic $W$ decays and
Higgs boson decays to $b\bar{b}$. There, the authors have found that
it is possible to discover a signal of the process at the $\sim 5 \sigma$
level at $\sim 100$~fb$^{-1}$ of luminosity, for points for which
$M_2 \sim 265-390$~GeV and $M_1 \sim 133-198$~GeV. Indeed, our analysis
is competitive with this result, with such points falling somewhere
between the 3$\sigma$ and 5$\sigma$ discovery regions at 100~fb$^{-1}$
and 300~fb$^{-1}$, as demonstrated
by the black and red curves respectively, in Fig.~\ref{fig:sig}. This indicates that this channel is as important as the final state with $h \rightarrow b\bar{b}$, or at least complementary.

\section{Conclusions}\label{sec:conclusions}
We have presented a phenomenological analysis of the channel $\Ntwo
\Cone \to (h \None) (W^\pm \None)$ using the $W \rightarrow \ell
\nu_\ell/ \tau \nu_\tau$ and Higgs boson
channels ($h\rightarrow \tau^+ \tau^-$ and $h \rightarrow W^+W^-
\rightarrow~\mathrm{leptons}$) at the LHC. Such channels are common in many
concrete SUSY models where the predictions include $\Ntwo$ and $\Cone$
that predominantly decay into $h$ and $W$, respectively. 

Our analysis has included detailed hadron-level simulation of the relevant dominant
backgrounds, including the effects of the underlying event. Hadronic $\tau$
identification was modelled at hadron level with a custom-made algorithm based on the ones employed by both
the ATLAS and CMS experiments. We have employed a cut-based analysis
on several variables that bring out the properties of the signal
against those of the backgrounds. Specifically, we have constructed a
mass variable, $M_{\mathrm{min}}$, which is sharply peaked at low value for the
$WZ$ background and broadly falls off for the signal. 

Consequently we have demonstrated the potential for discovering or constraining the SUSY
parameter space in the $M_2$-$M_1$ plane at integrated luminosities
of 100~fb$^{-1}$, 300~fb$^{-1}$ and 3000~fb$^{-1}$, collected at a 14~TeV proton-proton centre-of-mass
energy. 
The 5$\sigma$ discovery potential of our analysis reaches up to $M_2 \simeq 350$~GeV with $M_1 \lsim 100$~GeV
at the 14~TeV LHC with 300~fb$^{-1}$.
This implies that a future $e^+ e^-$ collider with $\sqrt{s} = 1$~TeV can play indispensable role to cover $M_2 < 500$~GeV region.
A large part of this region can also be covered by the 14~TeV High Luminosity LHC with $3000$~fb$^{-1}$,
which has a discovery potential in the $M_2 \lsim 550$~GeV, $M_1 \lsim 200$~GeV region. 

This work serves a first study of making use of $h \to \tau \tau$ mode in the chargino-neutralino searches. 
We thus recommend further examination of this channel by
experimental collaborations, including the effects of full detector
simulation, $\tau$-jet tagging and multi-variate analyses.

\appendix

\section{Definition of the \boldmath{$M_{\mathrm{min}}$} variable}\label{app:mmin}
\begin{figure}[!ht]
	\begin{center}
          \includegraphics[scale=0.50]{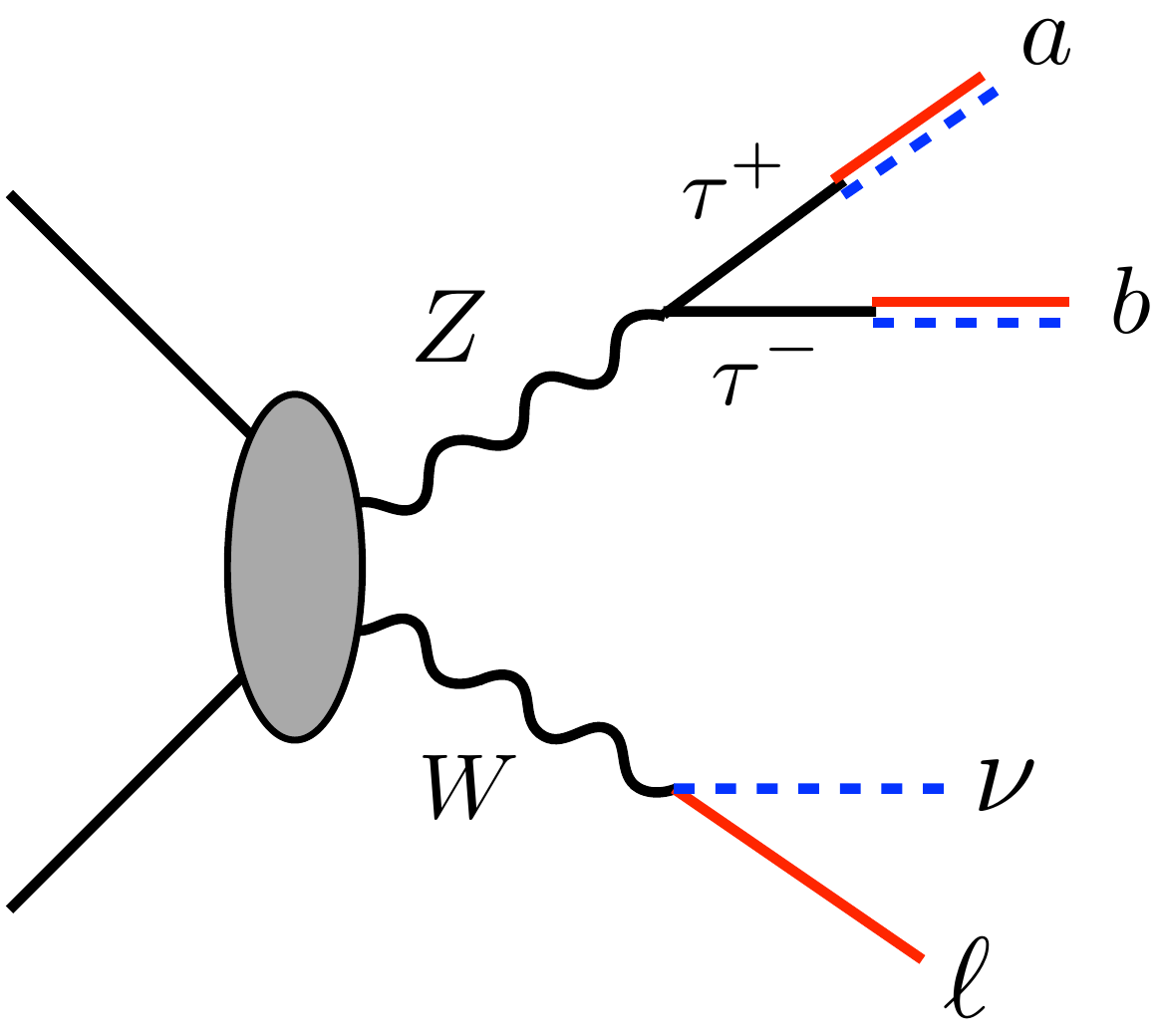}
	\caption{The $WZ$ background topology considered in
          constructing the $M_{\mathrm{min}}$ variable.}
	\label{fig:WZ_diagram}
	\end{center}
\end{figure}

We define the $M_{\rm min}$ variable that we will use as a handle for rejecting non-SUSY backgrounds. Although the variable is designed to
reject the $WZ$ background, it can also potentially perform well against
other backgrounds. There are three neutrinos in the final state: one coming from $W$
decay, the other two from  the $\tau$ lepton decays.
The direction of the $\tau$-neutrino is approximately collimated with
respect to the original $\tau$ lepton direction due to the mass hierarchy, $m_Z \gg m_{\tau}$. 
With this approximation, the momenta of the $\tau$ lepton and the $\tau$-neutrino can be parametrised as
\beqn
 p_{\tau^+} = { p}_{\rho_1}/a, &~~& { p}_{\tau^-} = { p}_{\rho_2}/b, \nonumber \\
 p_{\nu_1} = (1/a - 1) { p}_{\rho_1}, &~~& { p}_{\nu_2} =  (1/b - 1) { p}_{\rho_2},
\eeqn
where $p_{\rho_{1/2}}$ is the momentum of the visible decay
products and: $0<  a(b) < 1$. Note that events that in the
phenomenological analysis of this article that do not satisfy
this condition on $a$ and $b$ are deemed `unphysical' and rejected. 
Assuming the event topology in Fig.\,\ref{fig:WZ_diagram}, the unknown neutrino momenta can be constrained by
the mass shell conditions of the $W$ and $Z$ bosons and the missing momentum conditions.\footnote{Vectors in bold
  typeset represent 3-vectors.} 
\beqn
a, ~b, ~{\bf p}_{\nu}&:& ~5~{\rm unknowns}\nonumber \\
m_Z, ~m_W,  ~p_{\rm miss}^x, ~p_{\rm miss}^y &:& ~4~{\rm constraints} \nonumber 
\eeqn
Since ($\#$ of unknown $-$ $\#$ of constraints$)= 1$, we can parameterise the all neutrino momenta by a single parameter, $\theta$. 

The mass-shell constraint for the $Z$ boson gives
\beq
ab = \frac{2 ({p}_{\rho_1} \cdot {p}_{\rho_1}) }{ m_Z^2 }.
\eeq
By introducing $\theta \equiv \arctan \big( \frac{a}{b} \big)$, $a$ and $b$ can be written as
\beq
a = \sqrt{ \frac{2 ({ p}_{\rho_1} \cdot { p}_{\rho_1}) }{ m_Z^2 } \tan \theta }, ~~~~
b = \sqrt{ \frac{2 ({ p}_{\rho_1} \cdot { p}_{\rho_1}) }{ m_Z^2 } \tan^{-1} \theta }
\eeq
The transverse components of the neutrino momentum are determined by
\beq
{\bf p^T_\nu} = {\bf p_{\rm miss}^T} - (1/a - 1) {\bf p}_{\rho_1} - (1/b - 1) {\bf p}_{\rho_2}.
\eeq
The mass shell condition of $W$ constrains the last unknown parameter $p_\nu^z$ as
\beq
p_\nu^{z \pm} = \frac{ c p_\ell^z \pm \sqrt{E^2_\ell (c^2 - t_\ell^2 t_\nu^2 ) } }{  t_\ell^2 },
\label{eq:pz}
\eeq
where ${\bf t}_{\ell/\nu} = {\bf p}_{\ell/\nu}^T$, $c = {\bf t}_\ell \cdot {\bf t}_\nu + m_W^2/2$.
If Eq.~(\ref{eq:pz}) yields complex solution, we simply take the real
part~\cite{Gripaios:2010hv, Gripaios:2011jm}.

All the neutrino momenta are now parametrised by $\theta$.  We define the invariant mass of the system
\beq
M^\pm_{\rm inv}( \theta ) = \sqrt{ \big[ p_\ell + p^{\pm}_\nu(\theta) + p_{\tau^+}(\theta) + p_{\tau^-}(\theta) \big]^2 },
\eeq
where $\pm$ corresponds to the discrete ambiguity in Eq.~(\ref{eq:pz}).   
The variable $M_{\rm min}$ is defined by the global minimum of the $M_{\rm inv}$ over the $\theta$
\beq
M_{\rm min} \equiv \min_{\theta \in [0, \pi/2] } \min \{ M^+_{\rm inv}(\theta), M^-_{\rm inv}(\theta) \}  ~.
\eeq

\begin{figure}[!ht]
	\begin{center}
          \includegraphics[scale=0.50]{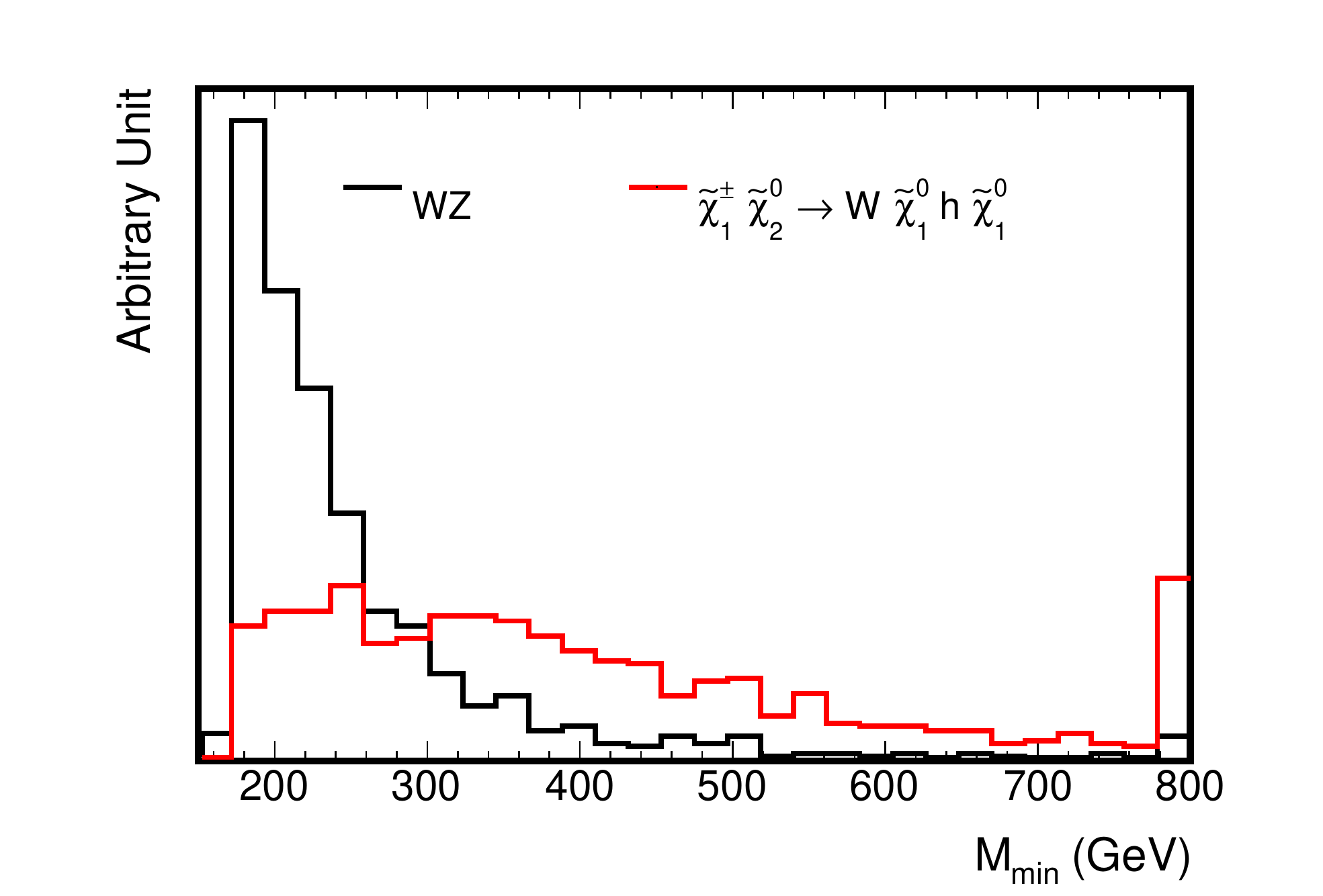}
	\caption{The $M_{\rm min}$ distribution for $WZ$ (black) and
          SUSY benchmark point (red) event samples.}
	\label{fig:Mmin_dist}
	\end{center}
\end{figure}

Fig.~\ref{fig:Mmin_dist} shows the distributions of $M_{\rm min}$ for
the $WZ$ and SUSY benchmark point event samples for 1000 parton-level events.  
The SUSY benchmark point \textbf{C350-100} involves the parameters: 
\beq
m_{\Cone} = m_{\Ntwo} = 350~{\rm GeV},~~~ m_{\None} = 100~{\rm GeV}.
\eeq

\section{Calculation of the initial cross sections}\label{app:crossx}
For completeness we provide the branching ratios used to reproduce the
initial cross sections that appear in Table~\ref{tb:benchanal}. 
\begin{itemize}
\item SUSY benchmark \textbf{C350-100}: Using
  \texttt{Prospino} 2.1, the NLO
  cross section for the SUSY benchmark point is $\sigma_{SUSY} \simeq
  200$~fb. For the signal, we consider the decays of the $W$ to all
  three lepton families and the
  decays of the Higgs boson to either $\tau^+\tau^-$ or
  $W^+W^-$ (again with the $W$s decaying to all leptons). We also assume
  that $BR(\tilde{\chi}_2^0 \rightarrow h \tilde{\chi}_1^0) = 1$. Hence:
\begin{eqnarray}  
\sigma(SUSY)_\mathrm{initial} &=& \sigma_{SUSY} \times BR(W
  \rightarrow \ell/\tau \nu)  \nonumber \\
&\times&
  ( BR(h\rightarrow \tau^+\tau^-) + BR(h\rightarrow W^+ W^-) \times
   BR(W
  \rightarrow \ell/\tau \nu)^2  )   \nonumber \\
  &=&\sigma_{SUSY} \times 0.3257  \nonumber \\
  &\times& ( 0.0632 + 0.2155 \times 0.3257^2) \nonumber  \\ 
  &\simeq& 5.7~\mathrm{fb}\;.
\end{eqnarray}
\item $WZ$: We allow for $(W\rightarrow \ell \nu, Z\rightarrow \tau^+
  \tau^-)$ or $(W\rightarrow \tau \nu, Z\rightarrow \ell^+
  \ell^-)$. We use the NLO cross section $\sigma(WZ) = 51.82$~pb,
  according to~\cite{Campbell:2011bn}. We obtain:
  $\sigma_\mathrm{initial} = \sigma(WZ) \times ( BR(W\rightarrow \ell
  \nu) BR(Z\rightarrow \tau^+ \tau^-) + BR(W\rightarrow \tau \nu) BR(Z
  \rightarrow \ell^+ \ell^-) ) = (51.82\times 10^3) \times ( 0.22
  \times 3.37 \times 10^{-2} + 0.11 \times (6.7 \times 10^{-2}) )~\mathrm{pb}
  \simeq 767$~fb. 
\item $W$+jets: The \texttt{AlpGen} tree-level cross section merged to
  the \textsf{HERWIG++} parton shower is $\sigma(W+\mathrm{jets})
  \simeq 300$~pb per lepton flavour (electrons, muons or taus). This was calculated for
  2 associated partons with the $W$ boson. 
\item $hV$: The processes $hW$ and $hZ$ are included. We assume $\sigma(hW) = 1504$~fb
  and $\sigma(hZ) = 883$~fb at NNLO QCD + NLO EW, taken
  from~\cite{Dittmaier:2011ti} for $M_h = 125$~GeV. We allow $h
  \rightarrow W^+W^-$ and $h\rightarrow \tau^+ \tau^-$, impose no
  constrain on the $W$ decays and allow for the $Z$ to decay to all
  leptons. Hence: $\sigma_\mathrm{initial} = (BR(h\rightarrow \tau^+\tau^-) + BR(h
  \rightarrow W^+ W^-)) \times  (\sigma(hZ)\times BR(Z\rightarrow
  \tau^+\tau^-/\ell^+ \ell^-) + \sigma(hW)) = (0.0632 + 0.2155) \times
  ( 883 \times (10.1\times10^{-2}) + 1504)~\mathrm{fb} \simeq 443$~fb.  
\item $t\bar{t}h$, $t\bar{t}$: We assume that the efficiency of
  tagging jets originating from the decays $b$ quarks is $70\%$. If
  one then vetoes events that contain at least one $b$-tagged jet,
  then for events containing $t\bar{t}$, a $1-(0.7^2 + 2 \times 0.3 \times
  0.7) = 0.09$ rejection factor can be
  achieved. We consider only leptonic decays of the $W$ bosons
  originating from the decays of the top quarks and only consider $h\rightarrow
  \tau^+ \tau^-$. We assume total cross sections: $\sigma(t\bar{t})
  \sim 900$~pb and $\sigma(t\bar{t}h) \simeq
  611$~fb~\cite{Dittmaier:2011ti}. This gives:
  $\sigma_\mathrm{initial}(t\bar{t} \rightarrow~\mathrm{leptons+jets})
  = 0.09 \times 900~\mathrm{pb} \times BR(W\rightarrow \ell/\tau \nu)^2
  \simeq 8600$~fb and $\sigma_\mathrm{initial}(t\bar{t}h
  \rightarrow~\mathrm{leptons+jets} +(\tau^+ \tau^-))
  \simeq 3.5$~fb. 
\item $Z$+jets: The \texttt{AlpGen} tree-level cross section merged to
  the \textsf{HERWIG++} parton shower is $\sigma(Z+\mathrm{jets})
  \simeq 300$~pb per lepton flavour (electrons, muons or taus). This sample has been produced with one associated
  parton with the $Z$ boson. 
\end{itemize}

\section{Discovery with low statistics}\label{app:statistics}
Discovery occurs when the probability of obtaining a given
experimental result, which contains some signal, is small when
compared to the expected background hypothesis. How small this
probability should be is somewhat a matter of preference and convention. Nowadays, in high energy physics, these probabilities are taken to
correspond to 3 standard deviations away from the assumed central
value of a Gaussian for the case of `evidence' of a
signal, and 5 standard deviations for the case of `discovery' of a
signal. On the other hand, exclusion is based on the probability of
having fewer events than the background alone would give, given the signal plus background
hypothesis.

\begin{figure}[t]
  \centering
  \begin{tabular}{cc}
    \includegraphics[width=0.4\linewidth]{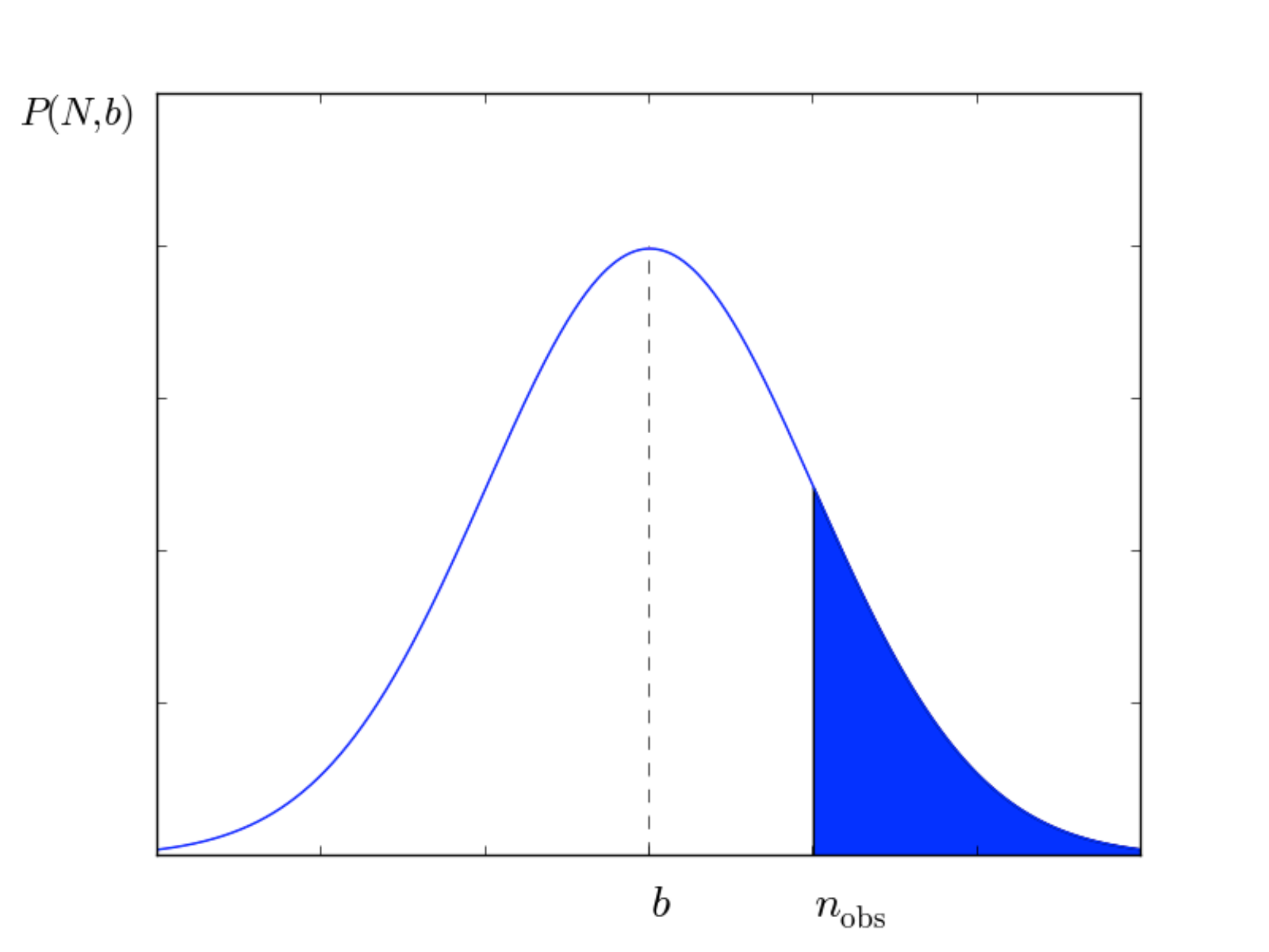}
   \end{tabular}
  \caption{The shaded region in the above probability distribution shows the
    probability of obtaining $N > n_{\mathrm{obs}}$ events.}
  \label{fig:shaded}
\end{figure}

To be concrete, let us assume that we are performing counting
experiments of events, obtaining as a result, $N_i$ counts
in each experiment $i$. Let us assume that in one specific experiment, we obtained
a measurement $n_{\mathrm{obs}}$. By some theoretical prediction, for
example obtained using a Monte Carlo event generator, or otherwise, the expected
background number of events in this experiment is given to be $b$. We
can assume that the counts $N_i$ are random variables, distributed
according to some distribution $P(N_i, b)$, where $b$ is the
mean of the distribution. In this case, the probability of obtaining
$n_{\mathrm{obs}}$ or more events, when the mean is equal to the expected background
$b$ is given by:
\begin{equation}\label{eq:probability}
P ( N \geq n_{\mathrm{obs}}, b) = \sum_{i=n_{\mathrm{obs}}}^{i = \infty}
P ( N_i, b), 
\end{equation}
where the sum can also be turned into an integral in the continuous
variable case. In simple words, according to the `background only'
distribution, getting a measurement of $n_{\mathrm{obs}}$ or more amounts to
the probability of the shaded area in Fig.~\ref{fig:shaded}., and this
probability tells you how likely $b$ is as an assumption of the mean of the
distribution. 

In the specific case of the Poisson distribution:
\begin{equation}
P\mathrm{ois}(N_i,b) = \frac{b^{N_i}}{N_i!} e^{-b},
\end{equation}
then Eq.~(\ref{eq:probability}) becomes $P ( N \geq n_{\mathrm{obs}}, b) = \sum_{i=n_{\mathrm{obs}}}^{i = \infty} \frac{b^{N_i}}{N_i!} e^{-b}$. This
sum can be shown (see Appendix~\ref{app:poisson}) to be equivalent to the so-called `regularised
incomplete gamma
function', $\Gamma_{\mathrm{reg}}(s,x)$:
\begin{equation}
P ( N \geq n_{\mathrm{obs}}, b)  = \sum_{i=n_{\mathrm{obs}}}^{i = \infty} \frac{b^N_i}{N_i!} e^{-b} =
\Gamma_{\mathrm{reg}}(n_{\mathrm{obs}},b) = \Gamma
(n_{\mathrm{obs}},b) / \Gamma(n_{\mathrm{obs}}),
\end{equation}
for $n_{\mathrm{obs}} > 0$, and where we have defined the `unregularised incomplete gamma function':
\begin{equation}
\Gamma(n_{\mathrm{obs}},b) = \int_0^b \mathrm{d} t~t^{n_{\mathrm{obs}}-1} e^{-t}\;,
\end{equation}
and $\Gamma(n_{\mathrm{obs}})$ is defined in Eq.~(\ref{eq:gamma}) in the
following section, for $n=n_{\mathrm{obs}}$.
We can then calculate the probability for discovery. This is given by $P ( N \geq n_{\mathrm{obs}}, b)$ for
$n_{\mathrm{obs}} = s+b$, where $s$ is the expected signal
contribution to the event counts. This probability will differ from
the one obtained using the large sample (i.e. Gaussian) approximation,
in which the significance is given approximately by $\sigma \sim
s/\sqrt{b}$. For exclusion, we need to calculate the probability of
having less than $b$ events, under the assumption that the expected
number of events is $s+b$, i.e. $P ( N < b, s+b)$.

\section{Cumulative distribution for Poisson
  random variables}\label{app:poisson}
The unregularised incomplete gamma function is given by:
\begin{equation}
\Gamma(n,x) = \int_0^x \mathrm{d} t~ t^{n-1} e^{-t}.
\end{equation}
One can then define the gamma function:
\begin{equation}\label{eq:gamma}
\Gamma(n) = \lim_{x\rightarrow \infty} \Gamma(n,x)= \int_0^\infty \mathrm{d} t~ t^{n-1} e^{-t}.
\end{equation}

The integral that appears in $\Gamma(n,x)$ can be expanded by
performing consecutive integrations by parts:
\begin{eqnarray}
\int_0^x \mathrm{d} t~ t^{n-1} e^{-t} &=& \left. -e^{-t} t^{n-1}
\right|^x_0 + (n-1) \int_0^x \mathrm{d} t~ t^{n-2} e^{-t} \nonumber \\
 &=&  -e^{-x} x^{n-1} - (n-1) e^{-x} x^{n-2} \nonumber \\
&&+\: (n-1) (n-2) \int_0^x \mathrm{d} t~ t^{n-3} e^{-t}\;, \nonumber \\
&=& -e^{-x} [ x^{n-1} + (n-1) x^{n-2} \nonumber \\
&& +\: (n-1) (n-2) x^{n-3} + ... ]\nonumber \\
&& +\: (n-1)(n-2)(n-3)...(1) [1 - e^{-x}]. \nonumber \\
\end{eqnarray}
From the last equality in the above expression we can deduce
that 
\begin{equation} 
\Gamma(n) = (n-1)!~.
\end{equation} 
For $n>0$, dividing $\Gamma(n,x)$ by $\Gamma(n)$, we obtain:
\begin{eqnarray}
\frac{\Gamma(n,x)}{\Gamma(n)} &=& 1 - e^{-x} \left[ \frac{x^{n-1}}{(n-1)!} +
  \frac{x^{n-2}}{(n-2)!} + \frac{x^{n-3}}{(n-3)!} + ... \right] \nonumber \\
&=& 1 - \sum_{i \leq n-1 } \frac{ x^i e^{-x} } { i! } = \sum_{i =n }^\infty \frac{ x^i e^{-x} } { i! } \;,
\end{eqnarray}
which is nothing but the cumulative sum for the Poisson distribution.

\section{Individual signal regions}\label{app:sr}

\begin{figure}[htb]
  \begin{center}
    \vspace*{1ex}
    \includegraphics[scale=0.33]{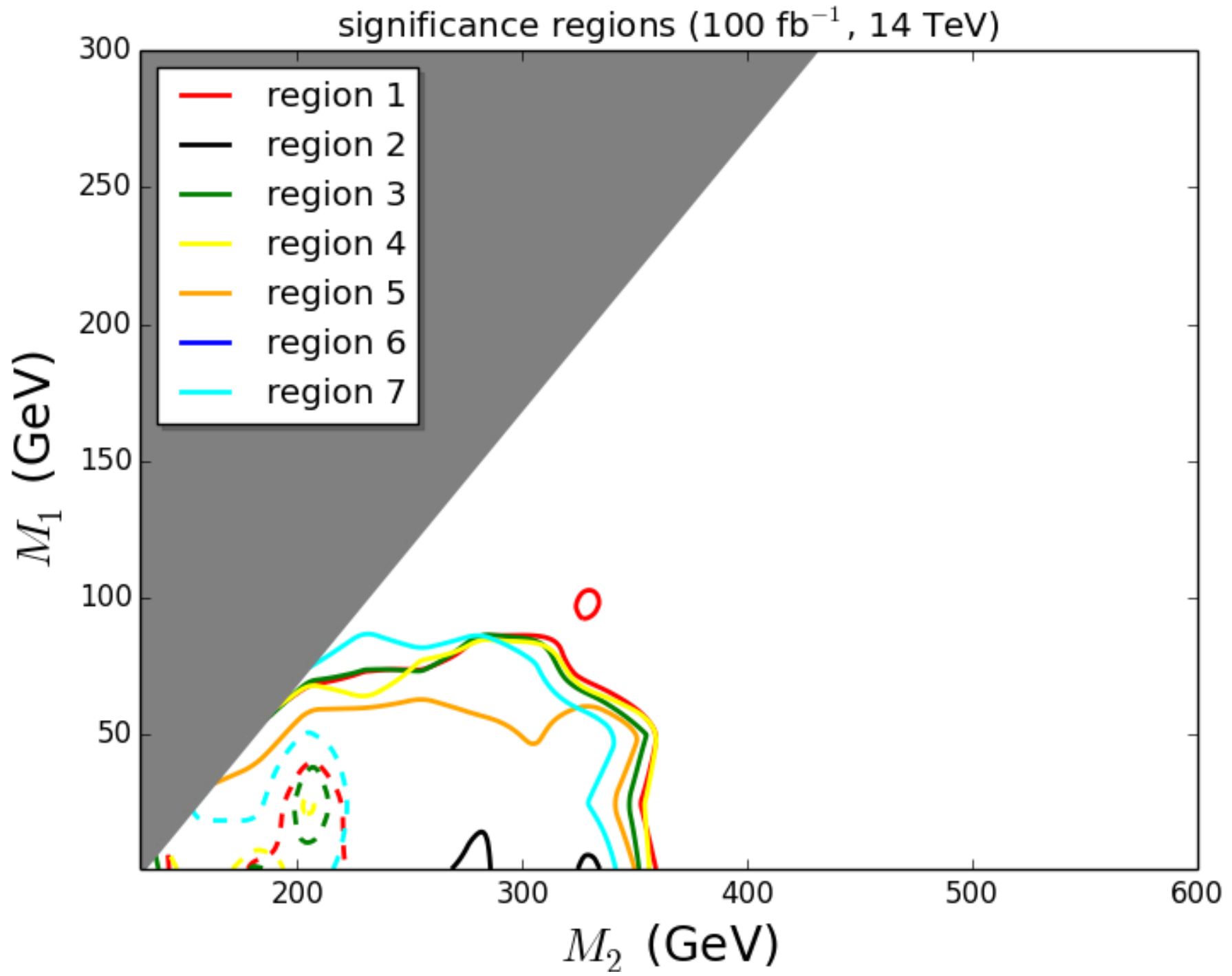}
    \includegraphics[scale=0.33]{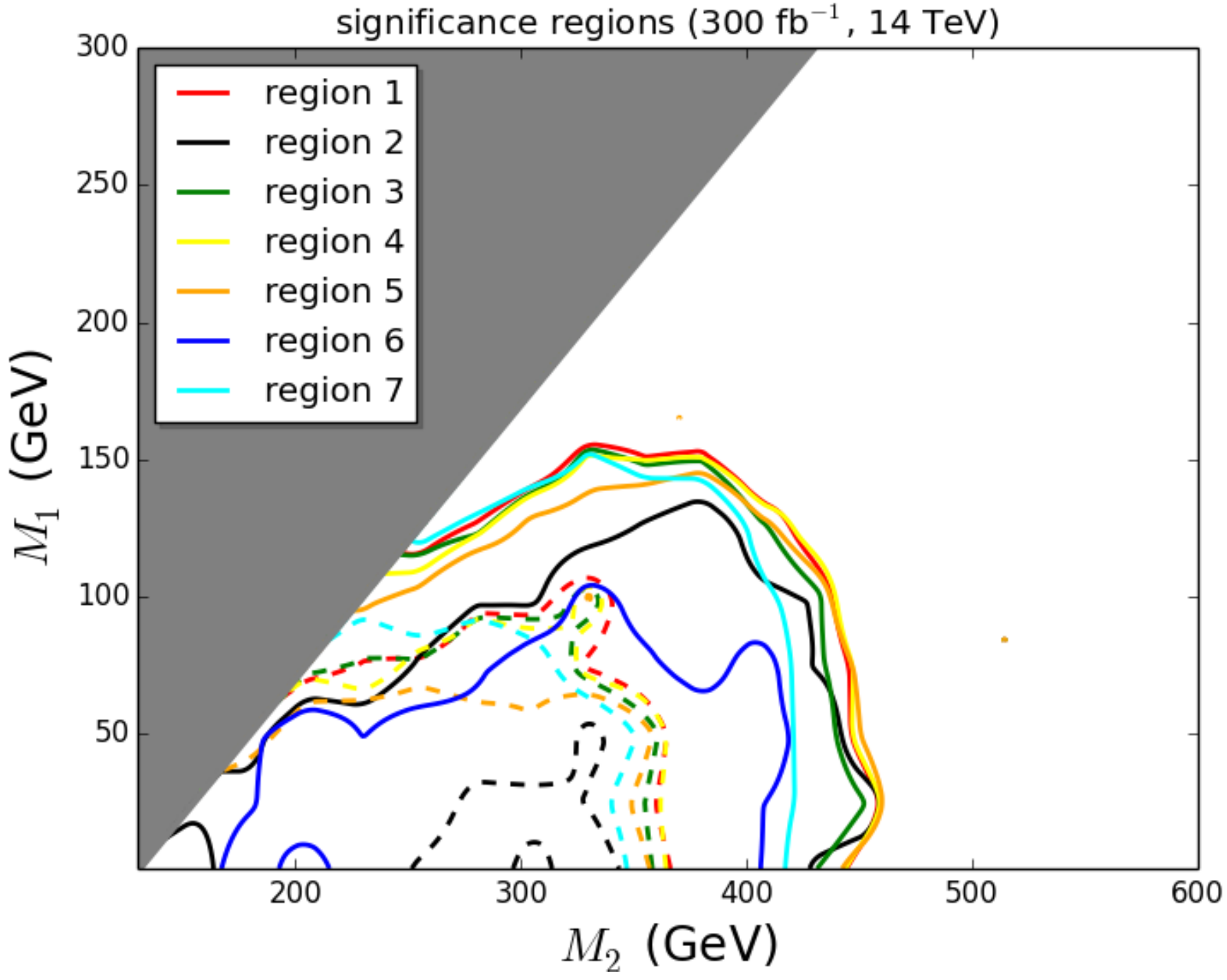}
    \includegraphics[scale=0.33]{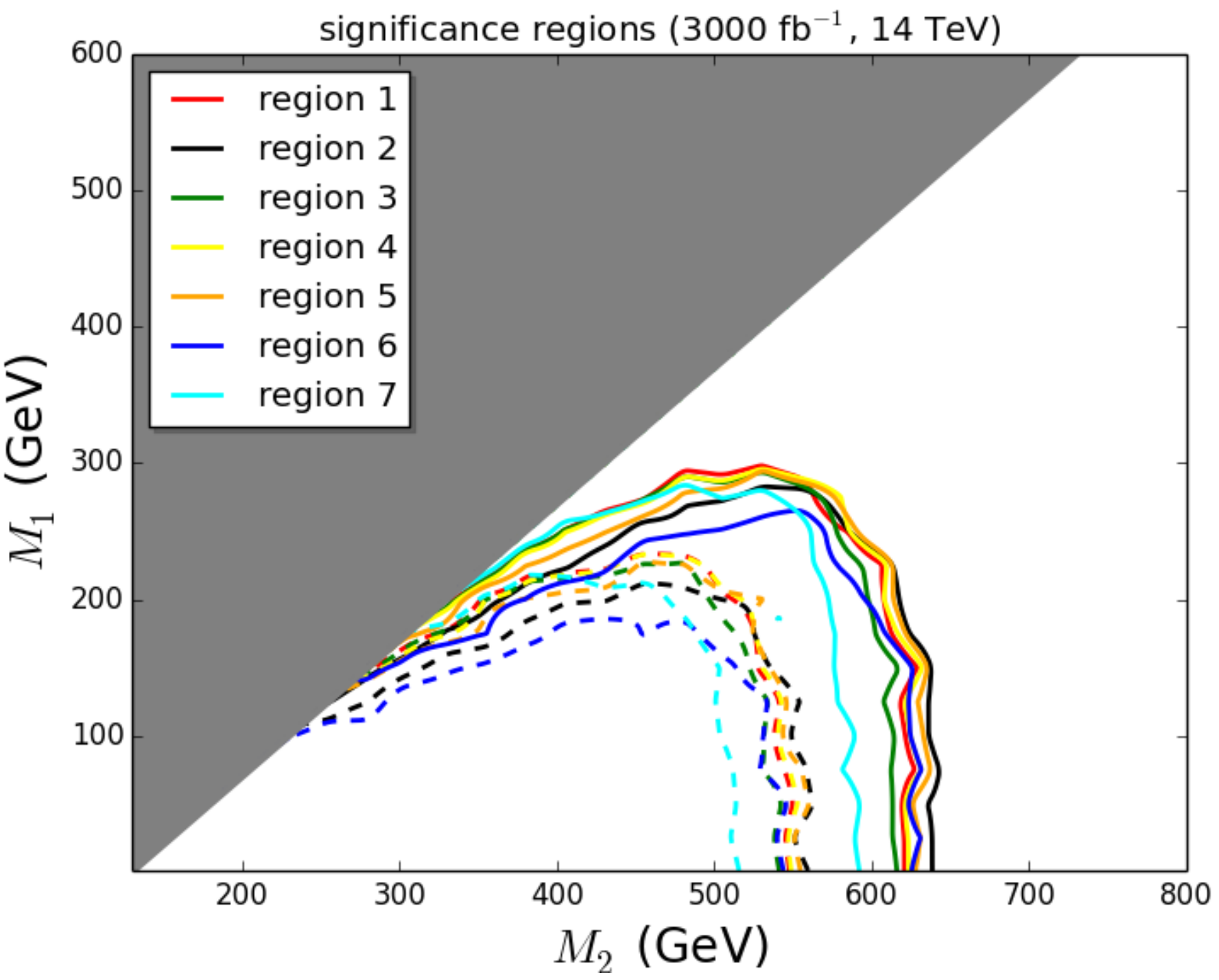}

  \end{center}
  \caption{The significance on the $M_2$-$M_1$ plane obtained for the
    signal regions defined in Table~\ref{tb:signalreg} at integrated
    luminosities of 100~fb$^{-1}$ (upper left), 300~fb$^{-1}$ (upper
    right) and 3000~fb$^{-1}$ (bottom). The solid curves show the $3\sigma$
   evidence region, whereas the dashed curves show the $5\sigma$ discovery region.}
  \label{fig:sigreg} 
\end{figure} 

\begin{figure}[htb]
  \begin{center}
    \vspace*{1ex}
    \includegraphics[scale=0.33]{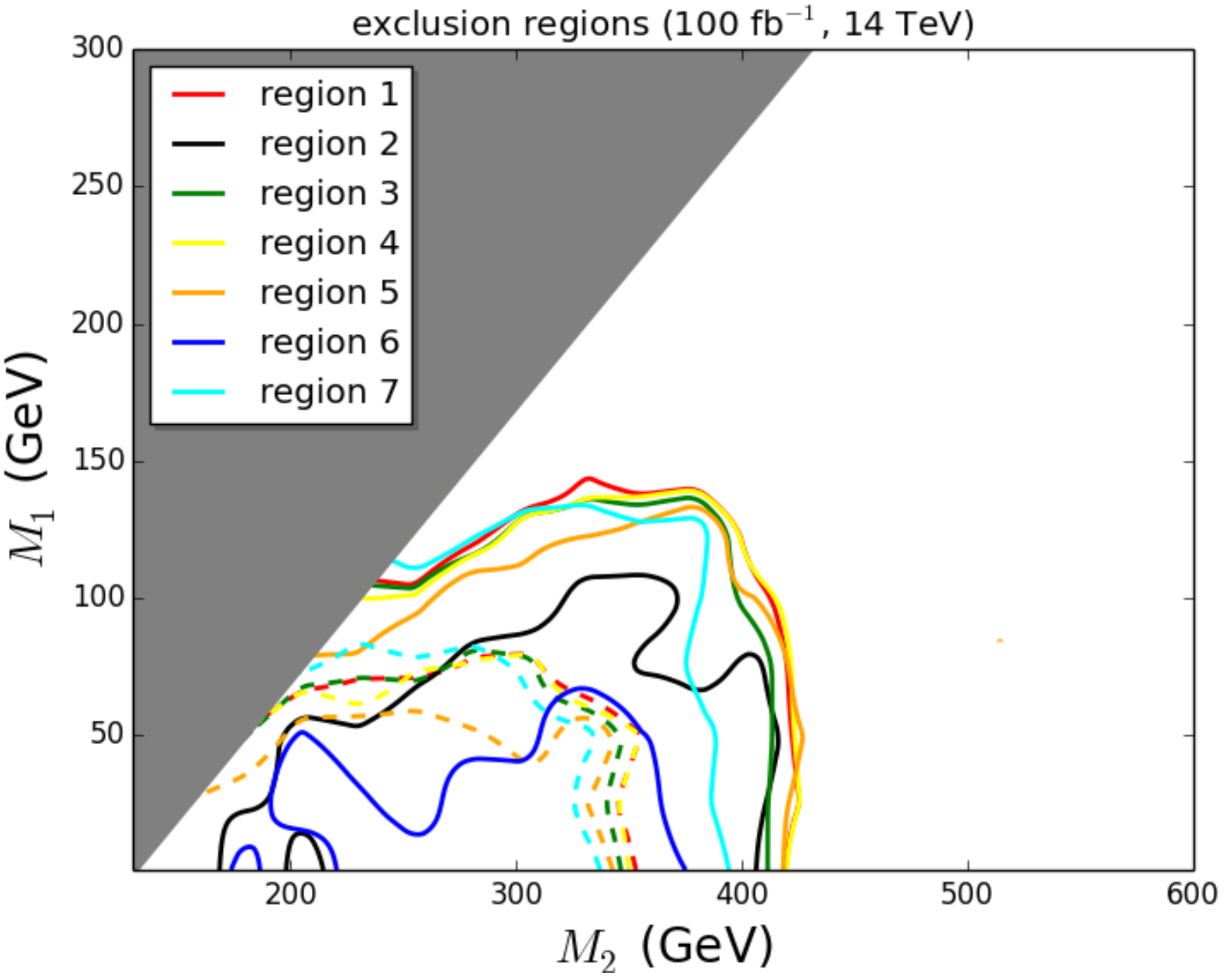}
    \includegraphics[scale=0.33]{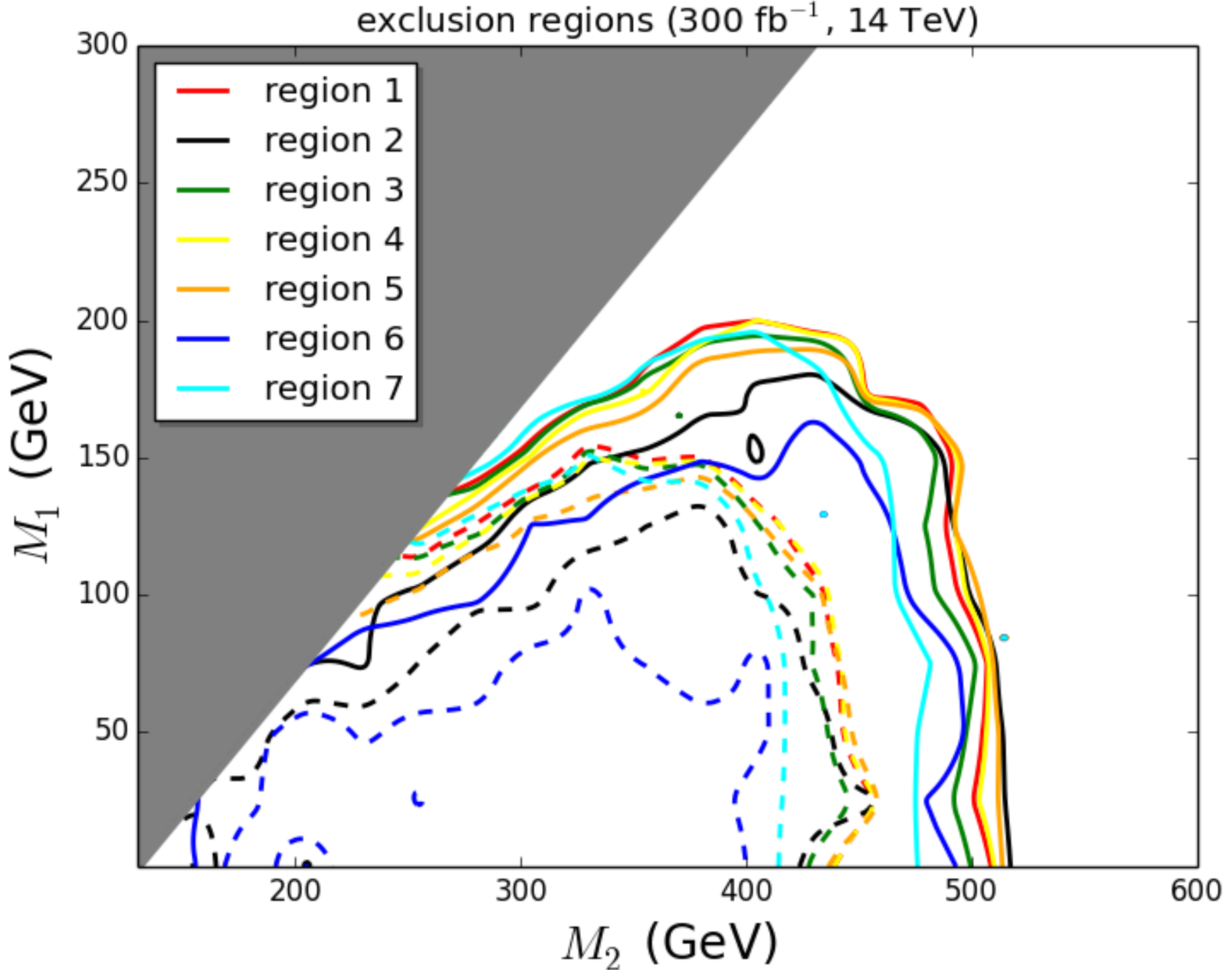}
    \includegraphics[scale=0.33]{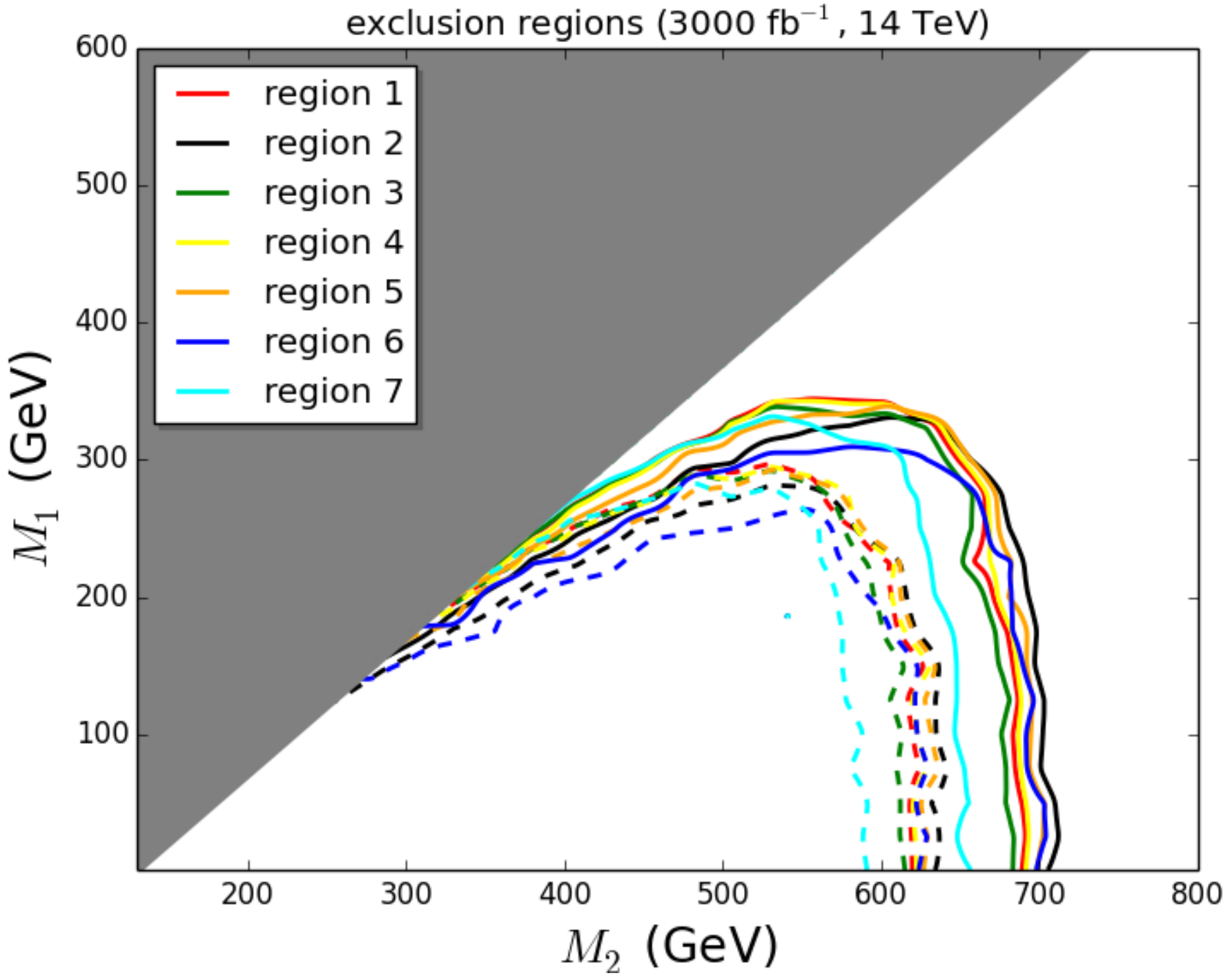}
  \end{center}
  \caption{The exclusion on the $M_2$-$M_1$ plane obtained for the
    signal regions defined in Table~\ref{tb:signalreg} at integrated
    luminosities of 100~fb$^{-1}$ (upper left), 300~fb$^{-1}$ (upper
    right) and 3000~fb$^{-1}$ (bottom). The solid curves show the $2\sigma$
   exclusion boundary, whereas the dashed curves show the $3\sigma$ boundary.}
  \label{fig:excreg}
\end{figure} 
In Figs.~\ref{fig:sigreg} and \ref{fig:excreg} we demonstrate the
individual signal regions contributing to the envelops shown in
Figs.~\ref{fig:sig} and \ref{fig:exc}. The analyses at each luminosity
are identical and the more `irregular' form at lower luminosities is
related to the Poisson statistics that govern the smaller number of
events in those cases.

In Table~\ref{tb:signalregbkg} we show the resulting cross sections
after applying each of the signal regions, defined in Table~\ref{tb:signalreg}. 
These can be used in conjunction with the efficiency data files for the signal on the
$M_2$-$M_1$ plane attached to this article\footnote{The file
  corresponding to signal region X is ``efficiency\_regionX\_expanded.dat'',
  located in the subdirectory ``efficiencies'' of the distribution of this article.}  to construct 
the signal cross section for each signal region for explicit BSM scenarios with 
$\tilde N \tilde C^{\pm} \to (h \chi) (W^\pm \chi)$ topology,
where $\tilde N$ and $\tilde C^{\pm}$ are massive BSM particles with the same mass, $M_2$, and $\chi$ is an invisible particle with mass $M_1$.
One can calculate the signal cross section for the process in question according to the
given model:
\beq
\mathrm{[signal~efficiency,~signal~region~X]} \times \mathrm{[signal~cross~section]} \times\mathrm{[BR]}\;,
\label{eq:sigcro}
\eeq
and use this in conjunction with the background cross section for
region X as given in the table to obtain the p-value over the
parameter space. 
Our efficiency data considers only the process with the $W \to \ell/\tau, \nu$ and Higgs bosons decaying inclusively to leptons (either
  $h\to \tau^+\tau^-$ or $h\to W^+W^- \to (e^+e^-, \mu^+ \mu^-, \tau^+\tau^-)+\slashed{E}_T$). 
The $[{\mathrm BR}]$ factor in Eq.~(\ref{eq:sigcro}) should therefore include these branching ratios.

\begin{table}[t!]
 \begin{center}
  \begin{tabularx}{\linewidth}{lXXXXXXX}
      \toprule
       signal region & SR1 & SR2 & SR3 & SR4 & SR5 & SR6 & SR7 \\ \midrule
       $\sigma_\mathrm{bkg}$ (fb) & 0.312 &  0.189 & 0.338 & 0.297 &
       0.254 & 0.195 & 0.512 \\ \bottomrule
    \end{tabularx}
 \end{center}
  \caption{The resulting sum of cross sections for the backgrounds for the different signal regions (SR) used in the analysis.}
\label{tb:signalregbkg}
\end{table}

\acknowledgments 
We would like to thank Jonas Lindert for interesting comments and
discussion. AP acknowledges support in part by the Swiss National Science Foundation (SNF) under contract 200020-149517, by
the European Commission through the ``LHCPhenoNet'' Initial Training
Network PITN-GA-2010-264564 and MCnetITN FP7 Marie Curie Initial Training Network
PITN-GA-2012-315877.
KS was supported in part by the London Centre for Terauniverse Studies (LCTS), using funding from
the European Research Council via the Advanced Investigator Grant 267352.
MT is grateful for funding from the Science and Technology Facilities Council (STFC).

\bibliography{susyhtautau.bib}
\bibliographystyle{JHEP.bst}

\end{document}